\documentclass[12pt]{article} 
\pdfoutput=1
\usepackage[utf8x]{inputenc} 
\usepackage[T1]{fontenc} 
\usepackage{amsmath,amssymb,bbm,commath,mathtools,wasysym,xfrac} 
\usepackage{graphicx} 
\graphicspath{{Figures/}}
\usepackage{color} 
\usepackage[unicode]{hyperref} 
\hypersetup{bookmarksnumbered=true, bookmarksopen=true, bookmarksopenlevel=2, breaklinks=true, citecolor=blue, colorlinks=true, linkcolor=red, linktoc=page, pdfborder={0 0 0}, pdfstartview=FitH, pdfauthor={PMCrichigno, DJain}, pdfsubject={Computing Nontoric Tri-Sasakian Volumes}, pdftitle={Non-toric Cones and Chern-Simons Quivers}, plainpages=false, unicode=true, urlcolor=cyan}
\usepackage{cite} 

\usepackage{geometry} 
\usepackage{fancyhdr} 
\usepackage{parskip} 
\usepackage[nottoc,notlof,notlot]{tocbibind} 
\usepackage[titles]{tocloft} 

\DeclareUnicodeCharacter{"0393}{\Gamma}
\DeclareUnicodeCharacter{"0394}{\Delta}
\DeclareUnicodeCharacter{"0398}{\Theta}
\DeclareUnicodeCharacter{"039B}{\Lambda}
\DeclareUnicodeCharacter{"039E}{\Xi}
\DeclareUnicodeCharacter{"03A0}{\Pi}
\DeclareUnicodeCharacter{"03A3}{\Sigma}
\DeclareUnicodeCharacter{"03A5}{\Upsilon}
\DeclareUnicodeCharacter{"03A6}{\Phi}
\DeclareUnicodeCharacter{"03A8}{\Psi}
\DeclareUnicodeCharacter{"03A9}{\Omega}
\DeclareUnicodeCharacter{"03B1}{\alpha}
\DeclareUnicodeCharacter{"03B2}{\beta}
\DeclareUnicodeCharacter{"03B3}{\gamma}
\DeclareUnicodeCharacter{"03B4}{\delta}
\DeclareUnicodeCharacter{"03B5}{\epsilon}
\DeclareUnicodeCharacter{"03B6}{\zeta}
\DeclareUnicodeCharacter{"03B7}{\eta}
\DeclareUnicodeCharacter{"03B8}{\theta}
\DeclareUnicodeCharacter{"03D1}{\vartheta}
\DeclareUnicodeCharacter{"03B9}{\iota}
\DeclareUnicodeCharacter{"03BA}{\kappa}
\DeclareUnicodeCharacter{"03BB}{\lambda}
\DeclareUnicodeCharacter{"03BC}{\mu}
\DeclareUnicodeCharacter{"03BD}{\nu}
\DeclareUnicodeCharacter{"03BE}{\xi}
\DeclareUnicodeCharacter{"03C0}{\pi}
\DeclareUnicodeCharacter{"03C1}{\rho}
\DeclareUnicodeCharacter{"03C3}{\sigma} 
\DeclareUnicodeCharacter{"03C4}{\tau}
\DeclareUnicodeCharacter{"03C5}{\upsilon}
\DeclareUnicodeCharacter{"03C6}{\phi}
\DeclareUnicodeCharacter{"03D5}{\varphi}
\DeclareUnicodeCharacter{"03C7}{\chi}
\DeclareUnicodeCharacter{"03C8}{\psi}
\DeclareUnicodeCharacter{"03C9}{\omega}
\DeclareUnicodeCharacter{"21D0}{\Leftarrow}
\DeclareUnicodeCharacter{"0212B}{\AA}
\DeclareUnicodeCharacter{"00B7}{\cdot}
\DeclareUnicodeCharacter{"266A}{\eighthnote}
\DeclareUnicodeCharacter{"266B}{\twonotes}
\PrerenderUnicode{ä}\PrerenderUnicode{ö}
\PrerenderUnicode{č}\PrerenderUnicode{³}

\definecolor{title}{rgb}{0.1,0.5,1.0}
\newcommand{\Title}[1] {\title{\color{title}\Huge #1}}
\newcommand{\TPheader}[3] {\thispagestyle{fancy}\pagenumbering{alph}\lhead{#1}\chead{#2}\rhead{#3}\cfoot{}}
\newcommand{\makepage}[1] {\newpage\pagenumbering{#1}}

\newcommand{\Abstract}[1] {\begin{abstract}\normalsize #1 \end{abstract}}

\newcommand\eqs[1] {\begin{align}#1\end{align}}
\newcommand\eqsn[1] {\begin{align*}#1\end{align*}}
\newcommand\eqss[1] {\begin{align}\begin{split}#1\end{split}\end{align}}
\newcommand\eqst[1] {\begin{multline}#1\end{multline}}
\newcommand\eqstn[1] {\begin{multline*}#1\end{multline*}}

\newcommand\eqsg[1] {\eqs{\begin{aligned}#1\end{aligned}}}
\newcommand\equ[1] {\begin{equation}#1\end{equation}}
\newcommand\equn[1] {\begin{equation*}#1\end{equation*}}

\newcommand\pmat[1] {\begin{pmatrix}#1\end{pmatrix}}

\renewcommand\exp[1] {e^{#1}}
\renewcommand\i {\dot{\iota}}
\newcommand\half {\tfrac{1}{2}}
\newcommand\s {\sigma}
\newcommand\ve {\varepsilon}
\renewcommand\( {\left(}
\renewcommand\) {\right)}
\newcommand\wh {\widehat}
\DeclareMathOperator{\Det}{Det}

\DeclareMathOperator{\tr}{tr}
\DeclareMathOperator{\Tr}{Tr}
\DeclareMathOperator{\Vol}{Vol}

\renewcommand\C {{\cal C}}
\newcommand\D {{\cal D}}

\newcommand\I {{\cal I}}
\newcommand\K {{\cal K}}
\renewcommand\L {{\cal L}}

\newcommand\N {{\cal N}}

\newcommand\bD {{\mathbb D}}
\newcommand\bH {{\mathbb H}}
\newcommand\bR {{\mathbb R}}

\newcommand\fm {\mathfrak{m}}
\newcommand\sT {{\sf T}} 
\newcommand\bs[1] {\boldsymbol{#1}}

\newcommand\nn {\nonumber\\}
\newcommand\VG {\Vol\(G\)}
\newcommand\Vsu {\Vol\(SU(2)\)}
\newcommand\Vu {\Vol\(U(1)\)}

\numberwithin{equation}{section} 
\begin{document}

\Title{Non-toric Cones and Chern-Simons Quivers}

\author{P. Marcos Crichigno$^{χ}$\footnote{\href{mailto:p.m.crichigno@uva.nl}{p.m.crichigno@uva.nl}}\,, 
Dharmesh Jain$^{ψ}$\footnote{\href{mailto:djain@phys.ntu.edu.tw}{djain@phys.ntu.edu.tw}}
\bigskip\\
\emph{${}^{χ}$Institute for Theoretical Physics, University of Amsterdam}\\ \emph{Science Park 904, Postbus 94485, 1090 GL, Amsterdam, The Netherlands}
\medskip\\
\emph{${}^{ψ}$Department of Physics, National Taiwan University}\\ \emph{No. 1, Sec. 4, Roosevelt Road, Taipei 10617, Taiwan}
}

\date{} 
\maketitle

\TPheader{\today}{}{} 

\Abstract{We obtain an integral formula for the volume of non-toric tri-Sasaki Einstein manifolds arising from nonabelian hyperk\"ahler quotients. The derivation is based on equivariant localization and generalizes existing formulas for  Abelian quotients, which lead to toric manifolds. The formula is particularly valuable in the context of AdS$_{4}\times Y_{7}$ vacua of M-theory and their field theory duals. As an application, we consider 3d $\N=3$ Chern-Simons theories with affine ADE quivers. While the $\wh A$ series corresponds to toric $Y_{7}$, the $\wh D$ and $\wh E$ series are non-toric. We compute the volumes of the corresponding seven-manifolds and compare to the prediction from supersymmetric localization in field theory, finding perfect agreement. This is the first test of an infinite number of non-toric AdS$_4$/CFT$_3$ dualities.
}

\makepage{Roman} 
\tableofcontents
\makepage{arabic}

\section{Introduction}

Sasaki-Einstein manifolds play an important role in AdS/CFT. These odd-dimensional manifolds, with the property that the cones over them are Calabi-Yau, appear naturally in the engineering of supersymmetric gauge theories by branes in string/M-theory. Their first appearance in holography was in the context of AdS$_5$/CFT$_4$. Placing $N$ D3-branes at the tip of a Calabi-Yau cone $\mathcal C(Y_5)$, and backreacting the branes, leads to an AdS$_5\times Y_5$ vacuum of Type IIB supergravity with a 4d $\N=1$ field theory dual. Following the first example of the conifold singularity $\C(T^{1,1})$ \cite{Klebanov:1998hh}, a vast number of new dualities were discovered by the explicit construction of an infinite family of Sasaki-Einstein metrics \cite{Gauntlett:2004yd}, and the subsequent identification of their field theory duals as quiver gauge theories \cite{Martelli:2004wu,Benvenuti:2004dy}.

Similar developments have followed in the case of AdS$_4$/CFT$_3$. Placing $N$ M2-branes at the tip of a hyperk\"ahler cone $\mathcal C(Y_7)$, where $Y_7$ is a {\it tri}-Sasaki-Einstein manifold now, and backreacting the branes leads to an AdS$_4\times Y_7$ vacuum of M-theory with a 3d $\N=3$ field theory dual. Following the first explicit example by ABJM \cite{Aharony:2008ug}, a large number of dual pairs have been identified, with $Y_{7}$ given by the base of certain hyperk\"ahler cones and the field theories corresponding to 3d $\N=3$ Chern-Simons (CS) quiver gauge theories \cite{Jafferis:2008qz,Herzog:2010hf,Gulotta:2011si,Jafferis:2011zi,Amariti:2011uw,Martelli:2011qj}.

Computing the volume of these manifolds is of great interest as the AdS/CFT dictionary relates $\Vol(Y)$ to important nonperturbative quantities in  field theory. For instance, in the case of D3-branes the $a$--anomaly coefficient of the 4d field theory is given by $a=\frac{\pi^3N^{2}}{4\Vol(Y_5)}$. In the case of M2-branes the free energy on the round three-sphere $F_{S^{3}}$ is given by  \cite{Drukker:2010nc,Herzog:2010hf}
\equ{\label{FS3vol}
F_{S^{3}}=N^{3/2}\sqrt{\frac{2\pi^{6}}{27\Vol(Y_{7})}}\,.
}
The independent evaluation of both sides of this relation has been crucial in providing convincing evidence for the proposed duality pairs. The LHS can be computed purely in field theory by supersymmetric localization \cite{Kapustin:2009kz} and has been carried out for a large number of CS quiver gauge theories \cite{Kapustin:2009kz,Herzog:2010hf,Drukker:2010nc,Gulotta:2011si,Jafferis:2011zi,Amariti:2011uw,Martelli:2011qj,Cheon:2011vi,Gulotta:2011aa,Crichigno:2012sk}. The RHS, however, has been mostly computed for {\it toric} $Y_{7}$,\footnote{A manifold $Y$ is toric tri-Sasaki Einstein if the cone $\mathcal C(Y)$ is a toric hyperk\"ahler manifold. A hyperk\"ahler manifold of quaternionic dimension $d$ is toric if it admits the action of $U(1)^d$ which is holomorphic with respect to all three complex structures. For a review of mathematical aspects of tri-Sasaki Einstein geometry, see \cite{Boyer:1998sf} and references therein.} and a detailed test of the duality for non-toric cases is lacking.\footnote{See \cite{Martelli:2009ga,Cheon:2011vi} for two non-toric examples, namely $V_{5,2}$ and $Q^{1,1,1}$.} The main reason for this is that although supersymmetric localization techniques are available on the field theory side for generic quivers, less tools are available on the geometry side for non-toric $Y_{7}$. 

The aim of this paper is to remedy this situation. Specifically, we provide a formula for computing the volumes of tri-Sasaki Einstein manifolds $Y_{4d-1}$ arising from nonabelian hyperk\"ahler quotients of the form $\C(Y_{4d-1})=\mathbb H^{d+∑_{a=1}^m n_a^2}///U(n_{1})\times\cdots \times U(n_{m})\,$. The derivation is based on the method of equivariant localization, making use of the $U(1)_{R}\subset SU(2)_{R}$ symmetry of the spaces. The localization method was developed in \cite{Witten:1982im,Moore:1997dj} and applied to toric hyperk\"ahler quotients, corresponding to the Abelian case, $n_{a}=1$,  by Yee in \cite{Yee:2006ba}. 

\begin{figure}[h!]
\centering
\includegraphics[scale=0.25]{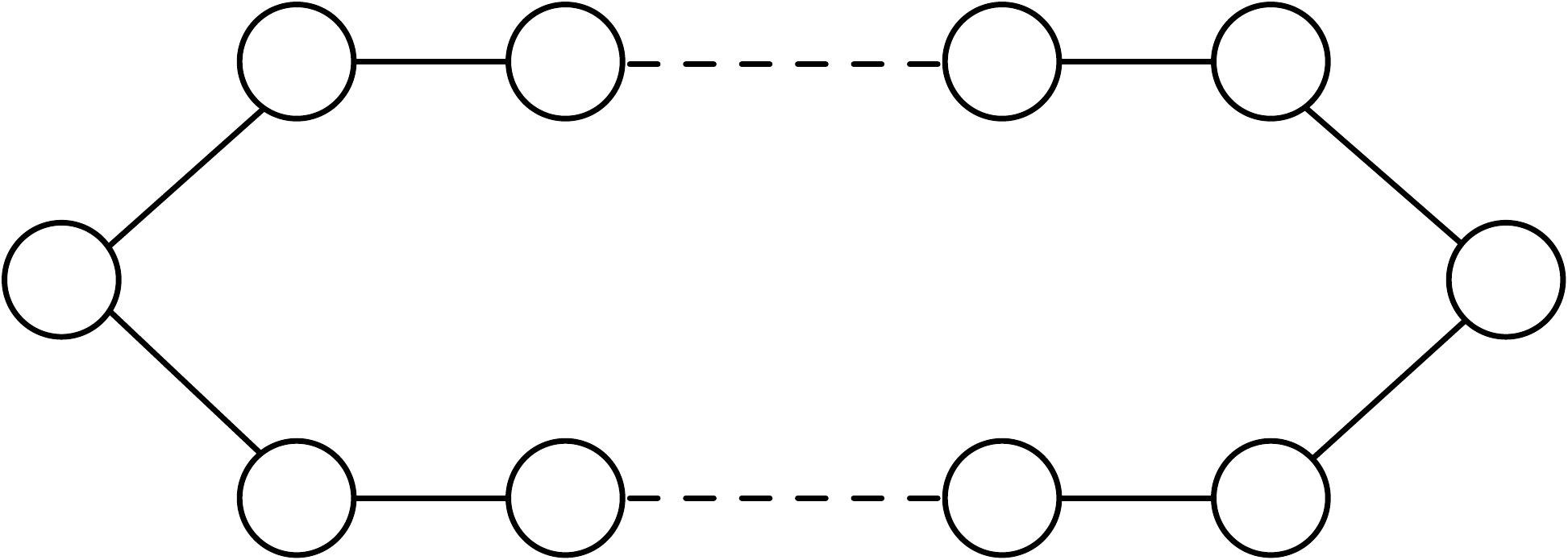} \qquad \includegraphics[scale=0.25]{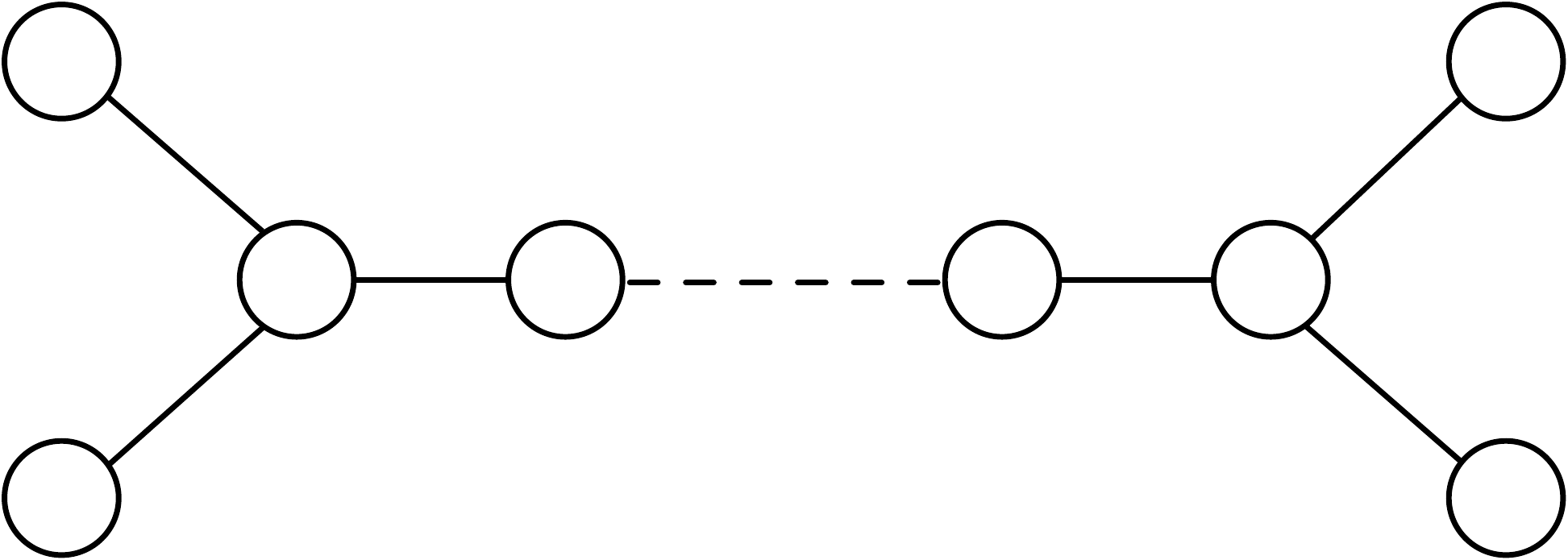} \\
\vspace*{5mm}
\includegraphics[scale=0.25]{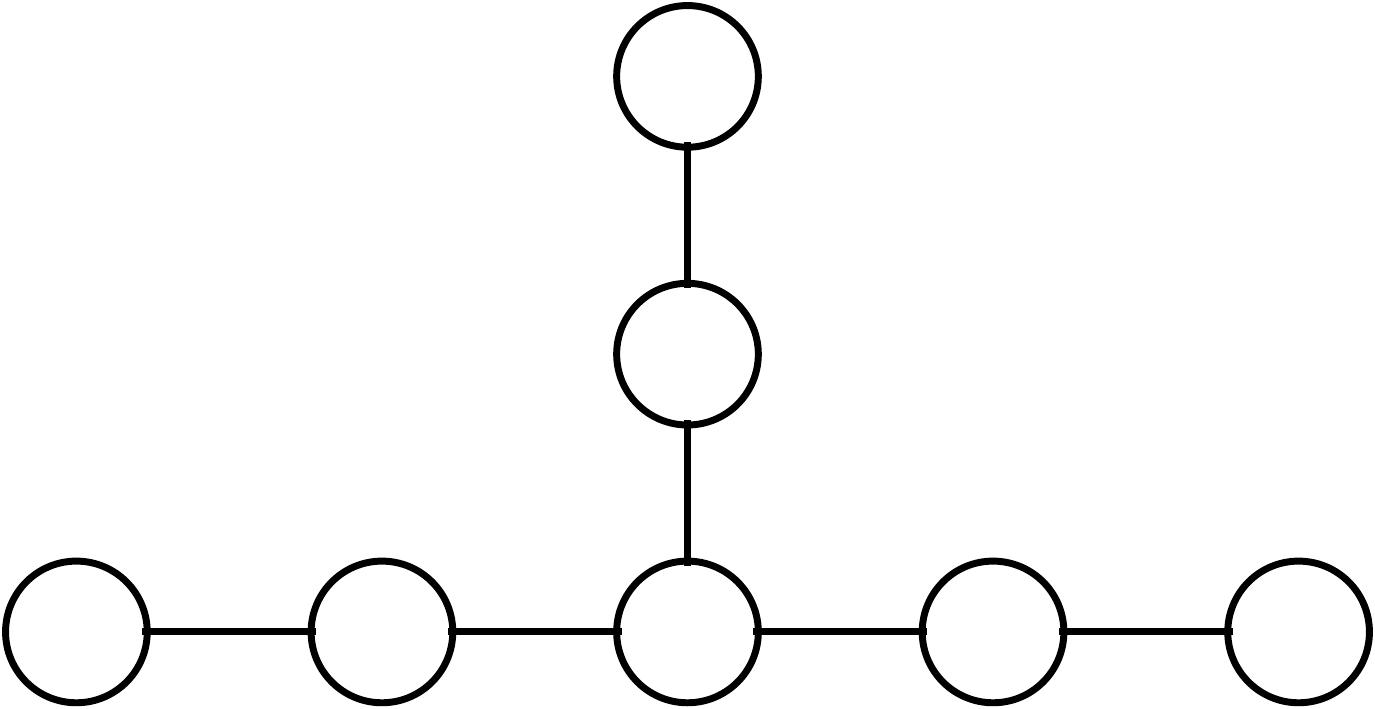} \quad \includegraphics[scale=0.25]{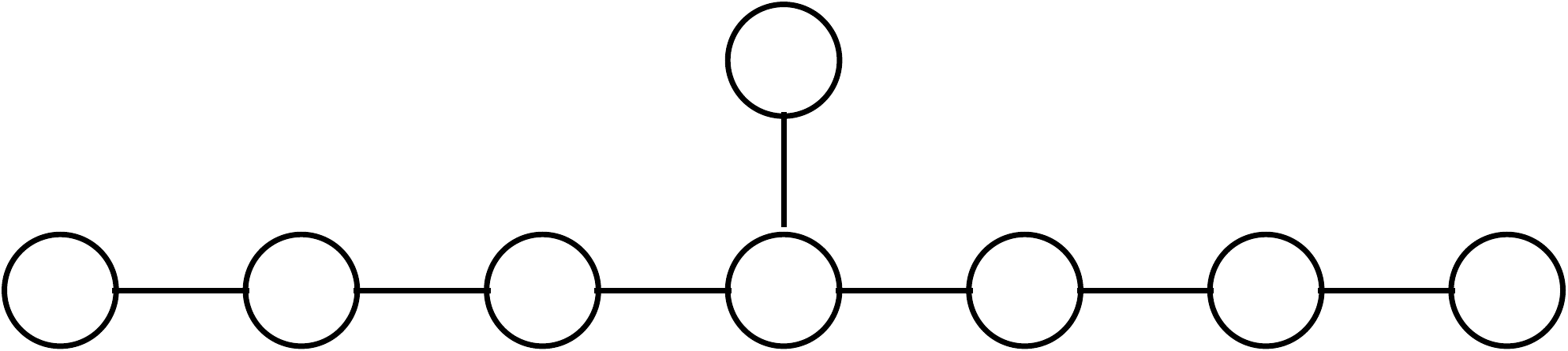} \quad \includegraphics[scale=0.25]{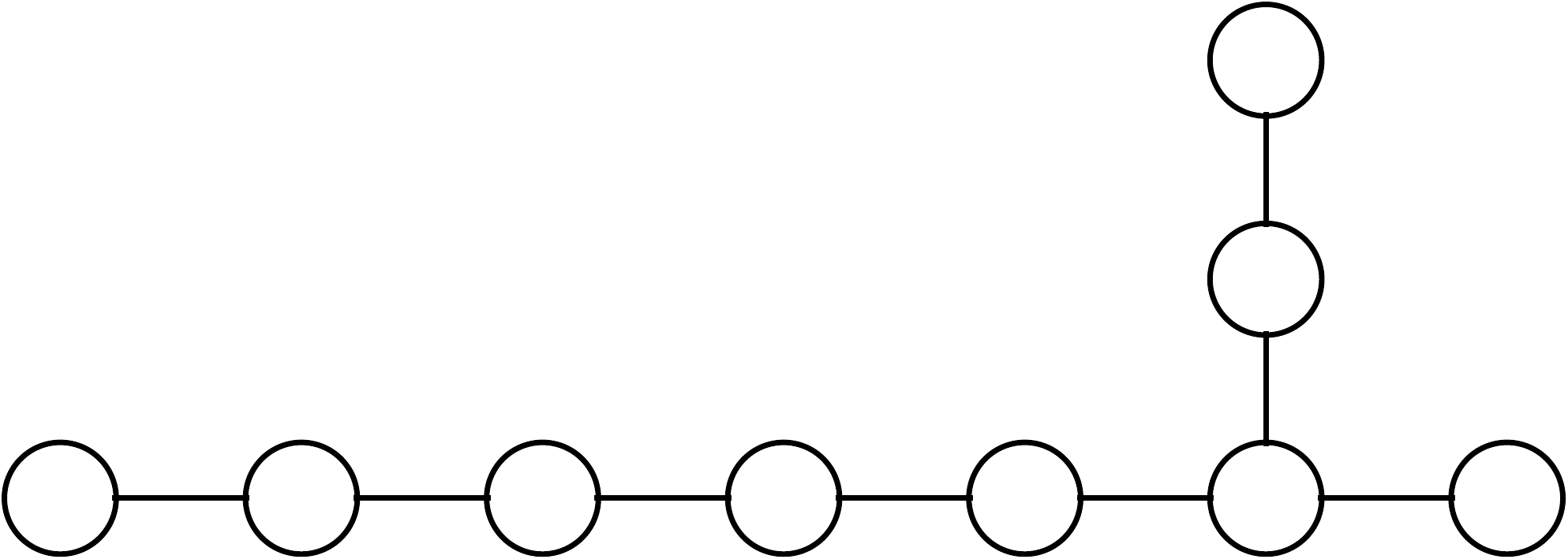}
\caption{Affine ADE quivers. From top to bottom and left to right $\wh A_{n},\wh D_{n}, \wh E_{6},\wh E_{7},\wh E_{8}$}
\label{Fig:quivers}
\end{figure}

Having derived a general formula, our main application is to 3d $\N=3$ CS matter quiver theories, whose field content is in one-to-one correspondence with extended ADE Dynkin diagrams -- see Figure~\ref{Fig:quivers}.
These theories \cite{Gulotta:2011vp} provide an ideal setting for applying the volume formula derived using localization. First, the corresponding tri-Sasaki Einstein manifolds can be constructed by hyperk\"ahler quotients and, while the $\wh A$ series is toric, the $\wh D$ and $\wh E$ series are non-toric. Second, as shown in  \cite{Gulotta:2011vp} for this class of field theories one may apply the saddle point approximation developed in \cite{Herzog:2010hf} to evaluate the free energy at large $N$. For the $\wh{A}$ series, both the evaluation of the free energy as well as the direct computation of the corresponding toric volume was carried out in \cite{Gulotta:2011si}, with perfect agreement. For the  $\wh D$ and $\wh E$ series, the free energy was computed by the authors  in \cite{Crichigno:2012sk}. In this paper we focus on the geometric side of the $\wh D$ series, identifying the precise tri-Sasaki Einstein manifolds and computing their volumes, finding perfect agreement with field theory. This is the first test of an infinite number of non-toric AdS$_4$/CFT$_3$ dualities. Few non-toric examples have been studied in the AdS$_{5}$/CFT$_{4}$ context; it is our hope that the formulas presented here will also be valuable in that context.

The paper is organized as follows. In the next section, we set up the localization procedure for computing the volumes of hyperk\"ahler quotients involving $U(N)$ or $SU(N)$ groups. Then, in section \ref{NThyperkählerQ} we specialize to $SU(2)^s×U(1)^r$ and provide a simple example. Finally, in section \ref{CSDQs} we study the moduli space of 3d $\N=3$ CS $\wh{D}$--quiver theories, identify the dual tri-Sasaki Einstein manifolds and compute their volumes. The volumes in the case of $\wh E$--quivers can also be computed by the techniques presented here, but we do not explicitly perform the corresponding integrals.

\section{Localization Setup}

In this section, we give a brief overview of the technical tools necessary for the computation of the volumes of hyperk\"ahler cones. The method was developed in \cite{Witten:1982im,Moore:1997dj} and is based on two basic features of the object we wish to compute. The first feature is the existence of a fermionic nilpotent symmetry of the symplectic volume integral, which allows one to localize the integral by adding an appropriate exact term. The second feature is that since these manifolds arise from hyperk\"ahler quotients of flat space, one may formulate the calculation in terms of the embedding flat space, where the calculations become simpler. We follow the exposition of Yee \cite{Yee:2006ba} (which we urge the reader to refer for more details), where this approach was applied to toric hyperk\"ahler quotients, and extend it to non-toric quotients.

Given a bosonic manifold $X$, and its tangent bundle $TX$ with canonical coordinates $\{x^\mu,V^\mu\}$, one defines the supermanifold $T[\psi]X$ obtained by replacing  the bosonic coordinates $\{V^\mu\}$ with fermionic ones $\{\psi^\mu\}$. Integrals of differential forms on $X$ can then be written as integrals of functions $f(x,\psi)$ over  $T[\psi]X$. For instance, the volume of a symplectic manifold $X$ with  symplectic 2-form $ω=\half ω_{μν}ψ^μψ^ν$ can be written as
\equ{\Vol(X)=∫_{T[ψ]X}e^ω\,;
\label{genvolxw}}
the Grassmann integration simply picks the correct power of $\omega$ to give the volume form on $X$. One may view this expression as a supersymmetric partition function; defining a `supersymmetry charge' $Q=ψ^μ\frac{∂}{∂x^μ}$ (which is  the de Rham differential, $d$), we see that the `action' $S=ω$ is supersymmetric, as $Qω=0$ (usually written as $dω=0$). Naïvely, one may want to use this nilpotent fermionic symmetry, $Q^2=0$,  to localize the integral. However, because $Q$ always contains a $\psi^\mu$, there is no $Q$--exact term one can add to the action which contains a purely bosonic term, required by the usual localization arguments. One way around this is to use a global symmetry of $\omega$ to deform $Q → {Q}_{\ve}$. Given a symmetry--generating vector field $V=V^μ\frac{∂}{∂x^μ}$ and defining the `contraction' by $V$ as $i_V=V^μ\frac{∂}{∂ψ^μ}$, there is a function $H$ such that $QH=i_Vω$, which can be named Hamiltonian, moment map, etc. depending on the context. This function $H$ can be used to deform the action to $S_\ve=ω -\ve H$, which is now invariant under ${Q}_{\ve}=ψ^μ\frac{∂}{∂x^μ}+\ve V^μ(x)\frac{∂}{∂ψ^μ}$. Moreover, ${Q}_{\ve}^2=\ve\L_{V}$ with the Lie derivative $\L_V=\{i_V,Q\}$, which implies that  ${Q}_{\ve}$ is nilpotent  in the subspace of $V$--invariant functions on $T[ψ]X$. This deformation now allows the addition of bosonic terms (with an $\ve$--dependence) and  localization can be performed. The next step is to combine this with the fact that the K\"ahler spaces of interest are obtained from a K\"ahler quotient of flat space.

\paragraph{Kähler Quotient.} Given a K\"ahler manifold $M$ with K\"ahler form $\omega$ and a holomorphic symmetry $G$, generated by  vector fields  $V_v$, $v=1,⋯,\dim G$, it follows  from $\mathcal L_{V_v}\omega=0$ that there are a set of moment map functions $μ_v$  satisfying $i_{V_v}ω=Qμ_v$. The Lie derivative $\L_{V_v}$ acts on the moment maps as follows
\equ{V_v^μ\frac{∂μ_u(x)}{∂x^μ}=i_{V_v}(Qμ_u)=i_{V_v}i_{V_u}ω=f_{uv}{}^wμ_w(x)\,,
\label{KLDonMM}}
where $f_{uv}{}^w$ are the structure constants of $G$. The submanifold $μ_v^{-1}(0)$ is $V$--invariant and the Kähler quotient $M//G$ is defined as the usual quotient $μ_v^{-1}(0)/G$. Parameterizing $M$ by splitting $\{x^μ\}$ into three parts $\{x^i,x^v,x^n\}$, such that $x^i \in μ^{-1}(0)/G$, $x^v$ denote the symmetry directions, i.e., $V_u=V_u^v\frac{∂}{∂x^v}$, and $x^n, n=1,⋯,\dim G$ are coordinates normal to $μ_v^{-1}(0)$, we can derive the following relations from $Qμ_v=i_{V_v}ω$:
\equ{∂_iμ_v=ω_{vi}\,,\qquad ∂_uμ_v=ω_{vu}\,,\qquad ∂_nμ_v=ω_{vn}\,.
}
Since $μ_v=0$ on $μ_v^{-1}(0)$, its derivative wrt $x^i$, $ω_{vi}=0$ on $μ_v^{-1}(0)$. Also, $ω_{vu}=0$ as $V_v^μ\frac{∂μ_u(x)}{∂x^μ}=0$ on $μ_v^{-1}(0)$. Thus, $Qω=0$ gives $∂_vω_{ij}=∂_i ω_{vj} -∂_j ω_{vi}=0$ so $ω_{ij}$ is $V$--invariant on $μ_v^{-1}(0)$ and the Kähler quotient then inherits $ω_{ij}$ as its Kähler form. Using \eqref{genvolxw}, the volume of the  quotient manifold can be written  as
\eqsn{\Vol\(M//G\) &=∫_{T[ψ]μ_v^{-1}(0)/G}[dx^i][dψ^i]\,e^{\frac{1}{2} ω_{ij}ψ^iψ^j} \\
&=\frac{1}{\VG}∫_{T[ψ]μ_v^{-1}(0)}[dx^v][dx^i][dψ^i]\,e^{\frac{1}{2} ω_{ij}ψ^iψ^j} \\
&=\frac{1}{\VG}∫_{T[ψ]M}[dx^n][dx^v][dx^i][dψ^i]\,e^{\frac{1}{2} ω_{ij}ψ^iψ^j}∏_{v=1}^{\dim G} δ\(μ_v(x)\)\left|\frac{∂μ_v(x)}{∂x^n}\right| \\
&=\frac{1}{(2π)^{\dim G}\VG}∫_{T[ψ]M}[dφ^v][dψ^v][dψ^n][dx^μ][dψ^i]\,e^{\frac{1}{2} ω_{ij}ψ^iψ^j}e^{\i φ^v μ_v +ψ^v ω_{vn} ψ^n}.
}
What these steps have achieved is to insert and exponentiate the moment map constraints to turn an integral over the quotient space $M//G$ into an integral over the embedding space $M$. Now, we use $ω_{vi}=ω_{vu}=0$ to write $ψ^v ω_{vn} ψ^n=ψ^vω_{vμ}ψ^μ$, where $μ$ runs over all values in $M$ (like $x^μ$). Next, inserting $ω_{in}$ and $ω_{mn}$ terms, which can be absorbed by shifting $ψ^i → ψ^i -ω_{ji}^{-1}ω_{jn}ψ^n$ and $ψ^v → ψ^v -ω_{nv}^{-1}ω_{nm}ψ^m$, to complete the $ω_{μν}ψ^μψ^ν$ term,  leads to the following simple expression:
\equ{\Vol\(M//G\) =\frac{1}{(2π)^{\dim G}\VG}∫_{T[ψ]M⊗φ^v}e^{ω+\i φ^vμ_v}.
}
One may further make use of the $U(1)_R$ symmetry to introduce the $\ve$--deformation
\equ{\Vol_{\ve}\(M//G\) =\frac{1}{(2π)^{\dim G}\VG}∫_{T[ψ]M⊗φ^v}e^{ω +\i φ^vμ_v -\ve H}\,
}
and  compute this integral by localization. When $M$ is multiple copies of the complex plane $\mathbb C$ with its canonical structures, the $\psi$--integrals are trivial and simply give 1. With appropriate $H$, the $x$--integrals are  Gaussian and only the integrals over $φ$'s remain, which require some more work to perform. The case of $M//G$  a conical Calabi-Yau six-fold is of interest for AdS$_{5}$/CFT$_{4}$. However, it should be emphasized  that the expression above computes the volume wrt the {\it quotient} metric, which is not necessarily (and typically is not) the Calabi-Yau metric on $M//G$.\footnote{One may consider, however, combining this with the principle of volume minimization \cite{Martelli:2005tp,Martelli:2006yb}. This should amount to performing the localization wrt a $U(1)_R'$ symmetry including possible mixings of $U(1)_R$ with flavor symmetries, but we do not study this here.} For this reason, we focus in what follows on hyperk\"ahler quotients, where the Calabi-Yau condition is automatic.

\paragraph{Hyperkähler Quotient.}
A hyperkähler manifold $M$ with a triplet of Kähler forms $\vec{ω}$ and a tri-holomorphic isometry group $G$ has triplets of moment maps satisfying $i_{V_v}\vec{ω}=Q\vec{μ}_v$. Most of what follows is a straightforward generalization of the Kähler case so we write down the most important equations only. The Lie derivative $\L_{V_v}$ acts on the moment maps as follows
\equ{V_v^μ\frac{∂\vec{μ}_u(x)}{∂x^μ}=i_{V_v}(Q\vec{μ}_u)=i_{V_v}i_{V_u}\vec{ω}=f_{uv}{}^w\vec{μ}_w(x)\,.
\label{LDonMM}}
The submanifold $\vec{μ}_v^{-1}(0)$ is $V$--invariant so the hyperkähler quotient $M///G$ is defined  \cite{Hitchin:1986ea} as the usual quotient $\vec{μ}_v^{-1}(0)/G$ . Parameterizing $M$ by $\{x^i,x^v,x^n\}$, where the only difference wrt the Kähler case is that $n=1,⋯,3\dim G$, we can derive from $Q\vec{μ}_v=i_{V_v}\vec{ω}$:
\equ{∂_i\vec{μ}_v=\vec{ω}_{vi}\,,\qquad ∂_u\vec{μ}_v=\vec{ω}_{vu}\,,\qquad ∂_n\vec{μ}_v=\vec{ω}_{vn}\,.
}
Again $\vec{ω}_{vi}=0$ and $\vec{ω}_{vu}=0$ on $\vec{μ}_v^{-1}(0)$. Thus, $Q\vec{ω}=0$ gives $∂_v\vec{ω}_{ij}=∂_i\vec{ω}_{vj} -∂_j\vec{ω}_{vi}=0$ so $\vec{ω}_{ij}$ is $V$--invariant on $\vec{μ}_v^{-1}(0)$ and the hyperkähler quotient then inherits $\vec{ω}_{ij}$ as its 3 Kähler forms. We pick $ω^3=ω$ to define the volume as
\eqsn{\Vol\(M///G\) &=∫_{T[ψ]\vec{μ}_v^{-1}(0)/G}[dx^i][dψ^i]\,e^{\frac{1}{2} ω_{ij}ψ^iψ^j} \\
&=\frac{1}{\VG}∫_{T[ψ]\vec{μ}_v^{-1}(0)}[dx^v][dx^i][dψ^i]\,e^{\frac{1}{2} ω_{ij}ψ^iψ^j} \\
&=\frac{1}{\VG}∫_{T[ψ]M}[dx^n][dx^v][dx^i][dψ^i]\,e^{\frac{1}{2} ω_{ij}ψ^iψ^j}∏_{v=1}^{\dim G}∏_{a=1}^3 δ\(μ^a_v(x)\)\left|\frac{∂μ^a_v(x)}{∂x^n}\right| \\
&=\frac{1}{(2π)^{3\dim G}\VG}∫_{T[ψ]M}[d\vec{φ}^v][d\vec{χ}^v][dψ^n][dx^μ][dψ^i]\,e^{\frac{1}{2} ω_{ij}ψ^iψ^j}e^{\i \vec{φ}^v·\vec{μ}_v +\vec{χ}^v·\vec{ω}_{vn}ψ^n}.
}
Again, these steps have turned an integral over $M///G$ to an integral over $M$. Now, using $\vec{ω}_{vi}=\vec{ω}_{vu}=0$ and relabelling $χ^v_3=ψ^v$ we rewrite $χ^v_3ω_{vn}ψ^n=ψ^vω_{vμ}ψ^μ$. Similarly, $χ^v_aω^a_{vn}ψ^n=χ^v_aQμ^a_v$, where $a=1,2$ now. Further relabelling $φ^v_3→φ^v$ and $φ^v_a→ρ^v_a$ and  inserting $ω_{in}$ and $ω_{mn}$ pieces, which can be absorbed by shifting $ψ$'s as before, one completes the $ω_{μν}ψ^μψ^ν$ term to  obtain a simplified exponent:
\equ{\Vol\(M///G\) =\frac{1}{(2π)^{3\dim G}\VG}∫_{T[ψ]M⊗φ^v⊗\{ρ^v_a,χ^v_a\}}e^{ω+\i φ^vμ_v +\i ρ^v_aμ^a_v +χ^v_aQμ^a_v}.
\label{volBloc}}

The `action' $S=ω+\i φ^vμ_v +\i ρ^v_aμ^a_v +χ^v_aQμ^a_v$ is invariant under a modified charge $\tilde{Q}$, acting on the `coordinates' as follows:
\eqsg{\tilde{Q}x^μ &= ψ^μ \\
\tilde{Q}ψ^μ &=-\i φ^vV_v^μ(x) \\
\tilde{Q}φ^v &=0 \\
\tilde{Q}χ^u_a &= -\i ρ^u_a \\
\tilde{Q}ρ^u_a &= -\i f_{vw}{}^uφ^vχ^w_a \,.
}
The transformation $\tilde{Q}ρ^u_a$ is fundamentally different from the toric case (where it vanishes), as a consequence of the  action of $\L_{V_v}$ on the moment maps \eqref{LDonMM}. However, it still squares as $\tilde{Q}^2=-\i φ^v\L_{V_v}$. Now we make use of the $U(1)_R\subset SU(2)_{R}$  symmetry to introduce the $\ve$--deformation and compute the integral by localization. This symmetry preserves only $ω^3=ω$, such that $i_Rω=QH$, and rotates the other two as $\L_R(ω^1-\i ω^2) = 2\i (ω^1-\i ω^2)$ $\big($also $\L_R(μ^1_v -\i μ^2_v) = 2\i (μ^1_v -\i μ^2_v)$ for all $v\big)$. The deformed action $S_{\ve}=S -\ve H$ is invariant under the deformed supercharge $\tilde{Q}_{\ve}$, which acts differently from $\tilde{Q}$ only on $ψ^μ$ and $ρ^u_a$, namely:
\eqsg{\tilde{Q}_{\ve}ψ^μ &=-\i φ^vV_v^μ(x) +\ve R^μ(x) \\
\tilde{Q}_{\ve}ρ^u_a &= -\i f_{vw}{}^uφ^v χ^w_a +2\ve ε_{ab}χ^u_b \,,
}
and  squares as $\tilde{Q}_{\ve}^2 = -\i φ^v\L_{V_v} +\ve\L_{R}$.

Now we are ready to localize  \eqref{volBloc} by adding the following term:\footnote{This useful trick is thanks to Kazuo Hosomichi.}
\equ{-t\tilde{Q}_{\ve}\big(\bar{x}^μ\tilde{Q}_{\ve}x_μ -χ^{+v}\tilde{Q}_{\ve}χ^{-v}\big) = -t\big(\bar{ψ}^μψ_μ +\bar{x}^μ\tilde{Q}_{\ve}^2 x_μ +ρ^{+v}ρ^{-v} +χ^{+v}\tilde{Q}_{\ve}^2χ^{-v}\big).
}
Here, $χ^{±}=(χ_1±\i χ_2)$ such that $\L_Rχ^-=2\i χ^-$ and the same for $ρ^±$. By taking the $t→+∞$ limit, the action $S_{\ve}$ does not contribute and the coordinates $x^μ, ψ^μ, ρ^v_a, χ^v_a$ can be simply integrated out, giving
\eqsn{&∫_{T[ψ]M⊗\{ρ^v_a,χ^v_a\}} e^{S_{\ve}-t\big(\bar{ψ}^μψ_μ +\bar{x}^μ\tilde{Q}_{\ve}^2 x_μ +ρ^{+v}ρ^{-v} +χ^{+v}\tilde{Q}_{\ve}^2χ^{-v}\big)} \\
=&\;(2t)^{\frac{\dim M}{2}}\(\frac{π}{t}\)^{\frac{\dim M}{2}}\frac{1}{\Det_M\tilde{Q}_{\ve}^2}\(\frac{π}{t}\)^{\dim G}(2t)^{\dim G}\Det_G\tilde{Q}_{\ve}^2 \\
=&\;(2π)^{\dim G+\frac{\dim M}{2}}\frac{\Det_G\tilde{Q}_{\ve}^2}{\Det_M\tilde{Q}_{\ve}^2}\,·
}
This leads to 
\equ{\Vol_{\ve}\(M///G\) =\frac{(2π)^{\dim G+\frac{\dim M}{2}}}{(2π)^{3\dim G}\VG}∫_{\{φ^v\}}\frac{\Det_G\tilde{Q}_{\ve}^2}{\Det_M\tilde{Q}_{\ve}^2}\,·
}
Here $\Det_G\tilde{Q}_{\ve}^2$ is simply the determinant of the $(\dim G)$--dimensional matrix $\(2\ve δ_w{}^u -f_{vw}{}^u φ^v\)$. $\Det_M\tilde{Q}_{\ve}^2$ depends explicitly on the manifold in consideration so we will tackle this in the next section.

For  $G=SU(2)$, $f_{uvw}=2ε_{uvw}$ and we can explicitly write the numerator in the above formula as
\equ{\label{volSU2}
\Vol_{\ve}\(M///SU(2)\) =\frac{(2π)^{3+\frac{\dim M}{2}}}{(2π)^9\Vsu}∫_{\vec{φ}}\frac{8\ve\big(\ve^2+\vec{φ}^2\big)}{\Det_M\tilde{Q}_{\ve}^2}\,·
}
This differs from the $U(1)$ case by the presence of $φ$'s in the numerator \cite{Yee:2006ba}:
\equ{\label{volU1}
\Vol_{\ve}\(M///U(1)\) =\frac{(2π)^{1+\frac{\dim M}{2}}}{(2π)^3\Vu}∫_{ϕ}\frac{2\ve}{\Det_M\tilde{Q}_{\ve}^2}\,·
}
We will distinguish the $U(1)$ variable by denoting it with $ϕ$ compared to $SU(2)$ variables $\vec{φ}$ from now on.

\section{Volumes of Non-toric Tri-Sasaki Einstein Manifolds}\label{NThyperkählerQ}

In this section, we consider the case of $G$ a product of multiple $SU(2)$'s and $U(1)$'s. At zero level the quotients will be the cones:
\equ{\label{GenDnQuotient}
\C\(Y^{(s,r)}_{4d-1}\)≡\bH^{d+3s+r}///SU(2)^s×U(1)^r\,.
}
As discussed in  detail in section~\ref{CSDQs}, these are the relevant quotients for $\wh D$--quiver CS theories.

We begin by setting up some notation. A quaternion $q$ can be written as 
\equ{\label{defquat}
q=\pmat{u & v \\ -\bar{v} & \bar{u}}
}
in terms of two complex variables $u$ and $v$. The flat metric is $ds^2=\frac{1}{2}\tr(dqd\bar{q})=dud\bar{u}+dvd\bar{v}$. The three Kähler forms are given by $\vec{ω}·\vec{\s}=\half dq\wedge d\bar{q}$:
\equ{ω^3=-\tfrac{\i}{2}\(du\wedge d\bar{u}+dv \wedge d\bar{v}\); \qquad (ω^1-\i ω^2)=\i(du\wedge dv)\,. 
}
Considering first $G=SU(2)×U(1)^r$, we realize the $SU(2)$ action on the quaternions $q$'s by pairing them up, i.e., we have $q^α_a$ with $α=1,2$ and $a=1,⋯,\frac{1}{2}\(d+3+r\)$. The quaternionic transformations are most simply given as:
\eqsg{δu^α_a &= u^β_a \left[\i (\vec{ζ}·\vec{\s})_β^α +\i {\textstyle ∑_{j=1}^r} Q_a^jξ_jδ_β^α\right] \\
δv^α_a &= -v^β_a \left[\i (\vec{ζ}·\vec{\s})_β^α +\i {\textstyle ∑_{j=1}^r} Q_a^jξ_jδ_β^α \right].
}
The vector fields corresponding to these symmetries are as follows:
\eqsg{V^r &=\tfrac{∂}{∂ξ_r}=\i{\textstyle ∑_a} Q_a^{r} \(u_a·∂_{u_a} -\bar{u}_a·\bar{∂}_{u_a} -v_a·∂_{v_a} +\bar{v}_a·\bar{∂}_{v_a}\) \\
V_3 &=\tfrac{∂}{∂ζ^3}=\i{\textstyle ∑_a} \(u^1_a∂_{u^1_a}-u^2_a∂_{u^2_a} -\bar{u}^1_a\bar{∂}_{u^1_a}+\bar{u}^2_a\bar{∂}_{u^2_a} -(u→v) \) \\
V_1 &=\tfrac{∂}{∂ζ^1}=\i{\textstyle ∑_a} \(u^2_a∂_{u^1_a} +u^1_a∂_{u^2_a} -\bar{u}^2_a\bar{∂}_{u^1_a} -\bar{u}^1_a\bar{∂}_{u^2_a} -(u→v) \) \\
V_2 &=\tfrac{∂}{∂ζ^2}=-{\textstyle ∑_a} \(u^2_a∂_{u^1_a} -u^1_a∂_{u^2_a} +\bar{u}^2_a\bar{∂}_{u^1_a} -\bar{u}^1_a\bar{∂}_{u^2_a} -(u→v) \).
}
Here `$·$' means sum over $α$.

Under the $SU(2)_{R}$ R-symmetry, each $q$ transforms by left action:
\equ{q→ e^{-\frac{\i}{2}\vec{\ve}·\vec{\s}}q\,,
}
such that the $U(1)_R\subset SU(2)_{R}$ symmetry  is generated by the  vector field
\equ{R=\i {\textstyle ∑_{a}}\(u_a·∂_{u_a} -\bar{u}_a·\bar{∂}_{u_a} +v_a·∂_{v_a} -\bar{v}_a·\bar{∂}_{v_a}\).
}
This  implies $i_R ω^3=QH$ with $H=\frac{1}{2}r^2 =\half ∑_{α,a}\(|u^α_a|^2+|v^α_a|^2\)$. It follows  that
\eqs{\Det_{q_a^α}\tilde{Q}_{\ve}^2 &=\left[\(\i \ve -{\textstyle ∑_{j=1}^r} Q_a^jϕ^j\)^2 -\vec{φ}^2\right]\left[\(\i \ve +{\textstyle ∑_{j=1}^r} Q_a^jϕ^j\)^2 -\vec{φ}^2\right] \nn
&=\left[\ve^2+\(|\vec{φ}|+{\textstyle ∑_{j=1}^r} Q_a^jϕ^j\)^2\right]\left[\ve^2+\(|\vec{φ}|-{\textstyle ∑_{j=1}^r} Q_a^jϕ^j\)^2\right].
\label{detqQsum}
}

For bifundamental quaternions wrt $G=U(2)_s×U(2)_{s+1}$, the transformations become $(\vec{τ}=\{I,\vec{\s}\})$:
\eqsg{δu^α_{aβ} &= u^γ_{aβ}\left[\i (\vec{ζ}_s·\vec{τ})^α_γ\right] -\left[\i (\vec{ζ}_{s+1}·\vec{τ})^γ_β\right]u^α_{aγ} \\
δv^α_{aβ} &= -v^γ_{aβ} \left[\i (\vec{ζ}_s·\vec{τ})^α_γ\right] +\left[\i (\vec{ζ}_{s+1}·\vec{τ})_β^γ\right]v^α_{aγ}\,.
}
This  leads to the following determinant (as per our convention, $φ^0≡ϕ$):
\eqs{\Det_{q^α_{aβ}}\tilde{Q}_{\ve}^2 &=\Big(\ve^2 +\big(|\vec{φ}_s|+|\vec{φ}_{s+1}| -(ϕ_s -ϕ_{s+1})\big)^2\Big)\Big(\ve^2 +\big(|\vec{φ}_s|+|\vec{φ}_{s+1}| +(ϕ_s -ϕ_{s+1})\big)^2\Big) \nn
&\quad\; \Big(\ve^2 +\big(|\vec{φ}_s|-|\vec{φ}_{s+1}| -(ϕ_s -ϕ_{s+1})\big)^2\Big)\Big(\ve^2 +\big(|\vec{φ}_s|-|\vec{φ}_{s+1}| +(ϕ_s -ϕ_{s+1})\big)^2\Big)\,.
}
For `bifundamentals' carrying more $U(1)$ charges, the $(ϕ_s -ϕ_{s+1})$ factor is simply replaced by a sum of all such charges $∑_iQ^i_aϕ^i$.

Thus, the (regularized) volumes of the hyperk\"ahler cones \eqref{GenDnQuotient} read:
\eqs{\Vol_{\ve}\Big(\C\(Y^{(s,r)}_{4d-1}\)\Big) &=\frac{(8\ve)^{s}(2\ve)^r (2π)^{3s+r+2(d+3s+r)}}{(2π)^{9s}(2π)^{3r}\Vol(SU(2)^s×U(1)^r)}∫_{\vec{φ}⊗ϕ}\frac{∏_{i=1}^s\!\!\big(\ve^2+\vec{φ}_i^2\big)}{\Det_M\tilde{Q}_{\ve}^2} \nn
&=\frac{2^{2d+3s+r}π^{2d}\ve^{s+r}}{\Vol(SU(2)^s×U(1)^r)}∫_{-∞}^∞∏_{i=1}^s d^3φ_i ∏_{j=1}^r dϕ_j \frac{∏_{i=1}^s\!\!\big(\ve^2+\vec{φ}_i^2\big)}{∏_{q\in M}\Det_q\tilde{Q}_{\ve}^2}\,·
\label{regvolC1}
}
To extract the volume of the tri-Sasaki Einstein base $Y$ from the $\ve$--regulated volume of the cone, recall that the conical metric is of the form $ds_{4d}^2=dr^2+r^2ds_{4d-1}^2$ and the $\ve H=\frac{\ve}{2}r^2$ term in $S_{\ve}$ serves as a regulator $e^{-\frac{\ve}{2}r^2}$ for the volume integral, giving the relation
\equ{\Vol_{\ve}\Big(\C\(Y^{(s,r)}_{4d-1}\)\Big) = \frac{2^{2d-1}Γ(2d)}{\ve^{2d}}\Vol\(Y^{(s,r)}_{4d-1}\).
\label{regVolC2}
}
Now, rescaling all $\{φ,ϕ\}→\{φ,ϕ\}/\ve$ in \eqref{regvolC1} to get rid of the factor $\ve^{3s+r}$ and comparing the result with \eqref{regVolC2} we obtain 
\equ{\frac{\Vol\(Y^{(s,r)}_{4d-1}\)}{\Vol(S^{4d-1})}=\frac{2^{3s+r}}{\Vol(SU(2)^s×U(1)^r)}∫∏_{i=1}^s d^3φ_i ∏_{j=1}^r dϕ_j\,\frac{∏_{i=1}^s\!\!\big(1+\vec{φ}_i^2\big)}{∏_{q\in M}(\Det_q\tilde{Q}_{\ve}^2)|_{\ve→1}}\,,
\label{intrepvol}}
where $\Vol(S^{4d-1})=\frac{2\pi^{2d}}{\Gamma(2d)}$. This is the main result obtained via the localization procedure. In section~\ref{CSDQs} we use this formula to compute the volume of tri-Sasaki Einstein manifolds relevant to 3d CS matter theories.   

\paragraph{General Quotients.} For a hyperk\"ahler quotient of the form $\mathbb H^{d+\dim G}///G$, the volume of the tri-Sasaki Einstein base is given by
\equ{\frac{\Vol\(Y_{4d-1}\)}{\Vol(S^{4d-1})}=\frac{1}{\VG}∫_{-∞}^∞∏_{i=1}^{\dim G}dφ^i\,\frac{\big|2δ_w{}^u -f_{vw}{}^uφ^v\big|}{(\Det_M\tilde{Q}_{\ve}^2)|_{\ve→1}}\,·
}
This integral over $\dim G$ $φ$'s can be reduced to $\mathop{\text{rank}} G$ $φ$'s in the `Cartan-Weyl basis', which introduces a Vandermonde determinant. For  $G$ a product of $U(N)$'s and (bi)fundamental quaternions we can write 
\equ{\frac{\Vol\(Y_{4d-1}\)}{\Vol(S^{4d-1})}=∫_{-∞}^∞∏_{U(N)∈G}\left[\frac{1}{N!}∏_{i=1}^{N}\frac{dλ^i}{2π}\right]\frac{{\displaystyle ∏_{U(N)∈G}}2^N{\displaystyle ∏_{i<j=1}^N}(λ_i -λ_j)^2\(4+(λ_i -λ_j)^2\)}{{\displaystyle∏_{\mathclap{\substack{i↔j \\ i∈U(M), j∈U(N)}}}}\(1+(λ_i -λ_j)^2\)}\,·
\label{genvolALE}}
We note that the factor $\VG$ has cancelled. When the quaternions are charged under more than two $U(1)$'s (as in $SU(M)×SU(N)×U(1)^r$), we need a change of basis to something similar to what we have for $SU(2)×U(1)^r$ in \eqref{detqQsum}. This can be achieved  by constraining the sum of eigenvalues of  $U(N)$ to vanish, reducing the number of variables to $(N-1)$, and adding a $ϕ$--variable for each $U(1)$ with the appropriate charge. The constant factors follow the same pattern as that for $U(N)$. Taking this into account, for a generic charge matrix $Q$ one obtains
\eqst{\frac{\Vol\(Y_{4d-1}\)}{\Vol(S^{4d-1})}=∫_{-∞}^∞∏_{SU(N)∈G}\left[\frac{1}{(N-1)!}∏_{i=1}^{N-1}\frac{dφ^i}{π}\right]∫∏_{U(1)∈G}\frac{dϕ}{π} \\
\frac{{\displaystyle ∏_{SU(N)∈G} ∏_{i<j=1}^N}(φ_i -φ_j)^2\(4+(φ_i -φ_j)^2\)}{{\displaystyle∏_{\mathclap{\substack{q_a \in \, i↔j \\ i∈SU(M),\,  j∈SU(N)}}}}\(1+(φ_i -φ_j -∑_{k}Q_a^kϕ^k)^2\)}\,,
\label{genvolSUNU1}}
where $φ_N=-∑_{i=1}^{N-1}φ_i$. This formula is applicable for generic quivers.

\subsection{An Example: ALE Instantons}\label{sec:ALE instantons}

As a simple example  we  consider four-dimensional ALE instantons. These are hyperk\"ahler quotients of the form $\mathbb H^{1+\dim G}///G$ with $G$ a product of unitary groups determined by an extended ADE Dynkin diagram \cite{kronheimer1989}. In the unresolved case, these spaces are simply cones over $S^{3}/\Gamma$  with $\Gamma$ a finite subgroup of $SU(2)$. The case $G=SU(2)^{k-3}\times U(1)^{k}$ with $k\geq 4$ corresponds to the $\wh D$ series and $\Gamma$ is the binary dihedral group $\mathbb D_{k-2}$ with order $4(k-2)$. This is precisely a quotient of the form \eqref{GenDnQuotient} so we may compute the volume of the base by the localization formula \eqref{intrepvol}. Let us work out the $k=4$ case first. Setting $d=1,s=1,r=4$, we have\footnote{Here we reduced the three-dimensional $SU(2)$ integral $\int_{-∞}^∞ d^{3}\phi$ to the obvious one-dimensional integral $\int_0^∞ d\phi(4π\phi^2)$. We recognize $φ^2$ as the `Vandermonde determinant'.}
\eqsn{\Vol\(Y^{(1,4)}_3\) &=\frac{2^8π^2}{π^2(2\pi)^4}∫_0^∞dφ(4π φ^2)\(1+φ^2\)∫_{-∞}^∞∏_{j=1}^4dϕ_j ∏_±\frac{1}{1+\big(φ±ϕ_j\big)^2} = \frac{π^2}{4}\,,
}
thus reproducing the expected volume $\frac{1}{8} \Vol\(S^3\)$.

For generic $k\geq 4$  we set $d=1,s=k-3,r=k$ in \eqref{intrepvol} and perform the integrals as in the example above. The computation is rather lengthy and thus we relegate the details to Appendix~\ref{BVolS3D}. The final answer is
\equn{
\Vol\(Y^{(k-3,k)}_3\) =\frac{2π^2}{4(k-2)}\,,
}
in accordance with the expected value of $\Vol\(\sfrac{S^3}{\mathbb D_{k-2}}\)$.

It is also possible to consider $\wh{E}_{6,7,8}$ singularities, corresponding to $G=U(3)×U(2)^3×U(1)^3$, $U(4)×U(3)^2×U(2)^3×U(1)^2$, and $U(6)×U(5)×U(4)^2×U(3)^2×U(2)^2×U(1)$, respectively. Using \eqref{genvolALE} or \eqref{genvolSUNU1} one obtains the expected volumes, given by $\Vol(S^3)$ divided by the order of tetrahedral (24), octahedral (48), and icosahedral (120) subgroups of $SU(2)$, respectively.

\subsection{Codimension 1 Cycles}

The volumes of codim-1 cycles are also of interest from the point of view of AdS/CFT correspondence, as they compute the conformal dimensions of chiral primary baryonic operators in the field theory. As discussed in \cite{Yee:2006ba}, a codim-1 cycle is defined by a holomorphic constraint that some $u=0$. This means that there are two types of such cycles for $\wh{D}$--quivers: $u_a^α=0$ or $u_{a,β}^α=0$. Let us focus on $u^1_1=0$ for concreteness but the computation does not depend on the explicit values of $a, α$. In the flat ambient space, this hypersurface is Poincaré dual to the 2-form
\equ{Γ_2 = δ(u_1^1)δ(\bar{u}_1^1)ψ^{u_1^1}\bar{ψ}^{\bar{u}_1^1},
}
with $QΓ_2=0=\tilde{Q}_{\ve}Γ_2$. This means the regularized volume of the $(4d-2)$--dimensional cone $u_1^1=0$ is simply obtained by
\equ{\langle Γ_2 \rangle_{\ve} =\frac{1}{(2π)^{3\dim G}\VG}∫_{T[ψ]M⊗φ^v⊗\{ρ^v_a,χ^v_a\}}Γ_2\,e^{ω+\i φ^vμ_v +\i ρ^v_aμ^a_v +χ^v_aQμ^a_v -\ve H}.
}
As the regularization is a simple Gaussian factor, this is related to the volume of $(4d-3)$--dimensional hypersurface inside the original cone by
\equ{\langle Γ_2 \rangle_{\ve} =\frac{2^{2d-2}Γ(2d-1)}{\ve^{2d-1}}\Vol\(Σ_{4d-3}\).
}
Evaluating the previous expression for $G=SU(2)^s×U(1)^r$ as before, the main difference is that the eigenvalue corresponding to $u_1^1$ is missing. Multiplying and dividing by it leads to
\eqst{\Vol\(Σ^{(s,r)}_{4d-3}\)=\frac{2^{3s+r+1}π^{2d-1}}{Γ(2d-1)\Vol(SU(2)^s×U(1)^r)}∫_{-∞}^∞∏_{i=1}^s d^3φ_i ∏_{j=1}^r dϕ_j\,∏_{i=1}^s\big(1+\vec{φ}_i^2\big) \\
\frac{1-\i\big(|\vec{φ}_1|+∑_jQ_1^jϕ^j\big)}{∏_{q∈M}(\Det_{q}\tilde{Q}_{\ve}^2)|_{\ve→1}},
\label{inthypvol}}
where the $\i Qϕ$ piece of the integrand vanishes because of the anti-symmetry under $ϕ→-ϕ$. The $φ_1$ piece can also be seen to vanish due to a cancellation from poles in the upper and lower half-planes. A similar numerator appears for the second type of cycle as well, for which we can take, as an example, $u_{5,1}^1=0$. Since the imaginary part of the integrand does not contribute, we obtain the same result as in the toric case, namely
\equ{\frac{\Vol\(Σ^{(s,r)}_{4d-3}\)}{\Vol\(Y^{(s,r)}_{4d-1}\)}=\frac{2d-1}{π}\,·
}

\section[\texorpdfstring{Chern-Simons $\wh{D}$--quivers}{ Chern-Simons D-quivers}]{Chern-Simons $\bs{\wh{D}}$--quivers}\label{CSDQs}

In this section, we consider the results of section~\ref{NThyperkählerQ} in the context of AdS$_{4}\times Y_{7}$ vacua of M-theory and their 3d field theory duals. Specifically, we are interested in CS $\wh D$--quivers,  whose gauge group is $U(2N)^{n-3}×U(N)^{4}$ with $n\geq 4$. The  main reason we focus on these theories is that it is a large class of theories for which the free energy has already been computed by supersymmetric localization \cite{Crichigno:2012sk} and the duals are non-toric.\footnote{The free energy of exceptional quivers was also computed in \cite{Crichigno:2012sk}. The computation of the corresponding volumes is straightforward (but tedious) with \eqref{genvolSUNU1} and the techniques developed in this section.} We begin by reviewing the  field theories.

\subsection{The Field Theories and their Free Energies}

\begin{figure}[h!]
\centering
\includegraphics[width=3in]{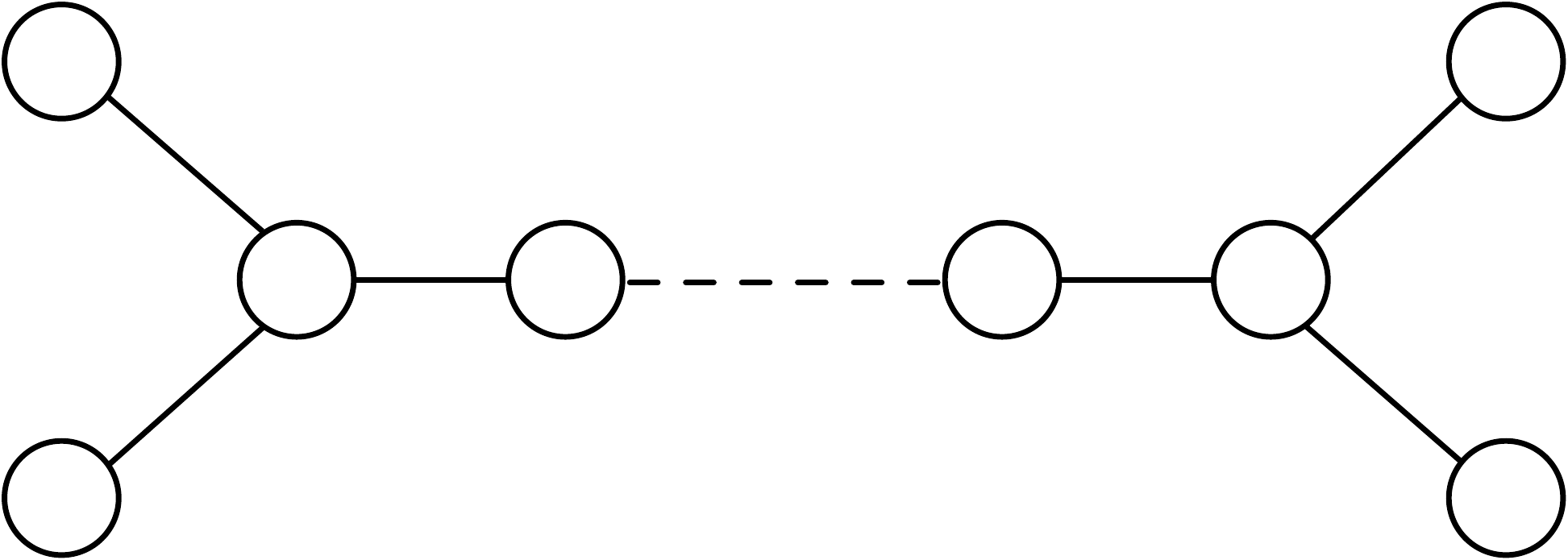}
\put(1,4){$k_3$} \put(1,65){$k_4$} \put(-228,65){$k_1$} \put(-228,4){$k_2$} \put(-180,50){$k_5$} \put(-142,50){$k_6$} \put(-115,50){$\cdots$} \put(-58,50){$k_{n+1}$}
\caption{$\wh D_n$ quiver diagram. Each node `$a$'  corresponds to a $U(n_a N)$ gauge group with CS level $k_a$, where $n_a$ is the node's comark and $\sum_a n_a k_a=0$ is imposed.}
\label{DnFig}
\end{figure}

The field content of the theories is summarized in the quiver of Figure~\ref{DnFig}. Following standard notation, we denote the fields in each edge of the quiver by $A,B$. We label the nodes and edges so that for a node $b>a$ the fields $A$ and $B$ associated to the edge $a\!\!↔\!\!b$ transform under $U(N_{a})\times U(N_{b})$ as $\bar{\mathbf N}_{a}\times \mathbf{ N}_{b}$ and $\mathbf{N}_{a}\times \bar{\mathbf N}_{b}$, respectively. The ranks of the gauge groups are given by  $N_a=n_aN$, with $n_a$  the node's comark  and the large $N$ limit corresponds to sending $N\to \infty$ (and CS levels fixed). The labelling of the nodes and their corresponding CS levels are shown in Figure~\ref{DnFig}.

With these conventions the action is given by 
\eqs{\label{action}S=S_{CS}&+\int d^4\theta\,\Bigg[\sum_{i=1}^2 \({A_i}^{\dagger}e^{V_5}A_{i}e^{-V_i}+B_{i}^{\dagger}e^{V_{i}}B_ie^{-V_5}\)+\sum_{i=3}^4 \({A_i}^{\dagger}e^{V_{n+1}}A_{i}e^{-V_i}+B_{i}^{\dagger}e^{V_{i}}B_ie^{-V_{n+1}}\)\nn
&+  \sum_{i=5}^{n} \({A_i}^{\dagger}e^{V_{i+1}}A_{i}e^{-V_i}+B_{i}^{\dagger}e^{V_{i}}B_ie^{-V_{i+1}}\)\Bigg] +\(\int d^{2}\theta\,W+\text{h.c.}\),
}
where $S_{CS}$ is the standard supersymmetric CS action (see e.g. \cite{Aharony:2008ug} and references therein) and $W$ is a superpotential term, which we will write explicitly below.  

\begin{figure}[h!]
\centering
\includegraphics[scale=0.75]{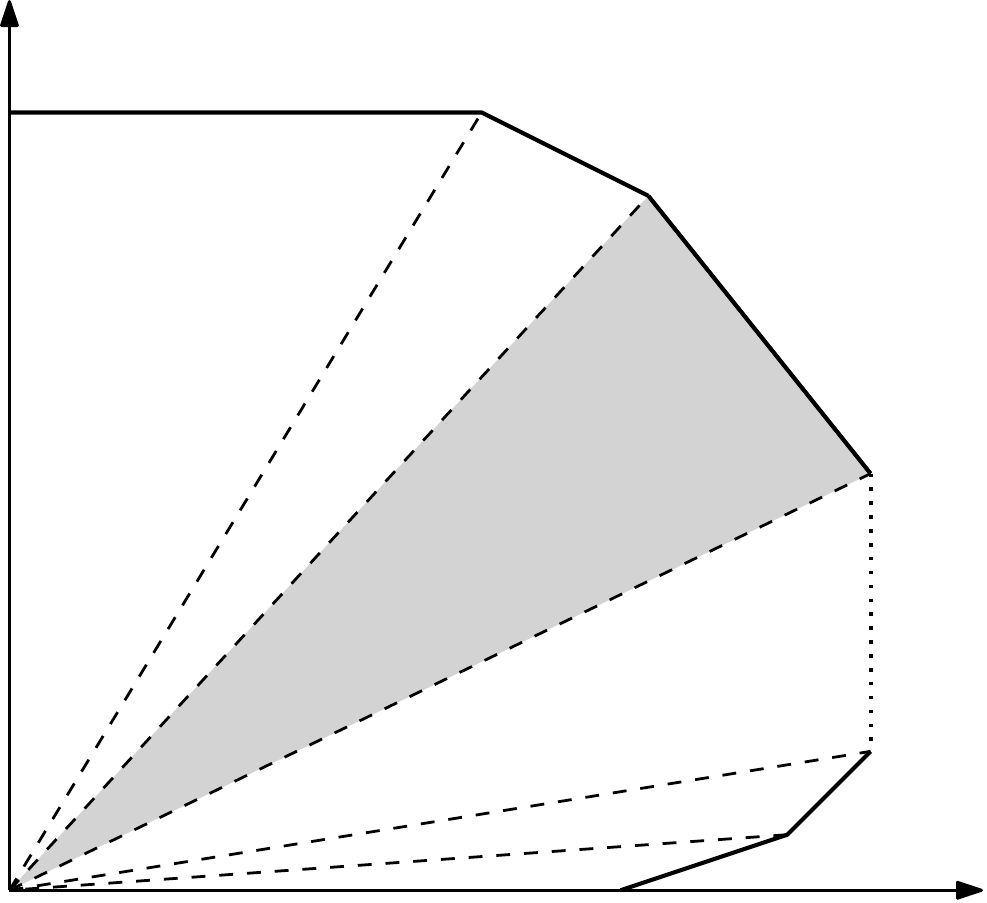}
\put(-70,155){$a$} \put(-20,95){$a+1$} \put(-90,105){$\frac{1}{2}\frac{\gamma_{a,a+1}}{\bar \sigma_a\, \bar \sigma_{a+1}}$} \put(-225,190){$y$} \put(5,0){$x$}
\caption{Schematic form of the polygon $\mathcal P_n$ for the $\widehat D_n$  quiver for a generic value of CS levels. Only the upper right quadrant is shown as it is symmetric along both the $x$ and $y$ axes.}
\label{Cone}
\end{figure}
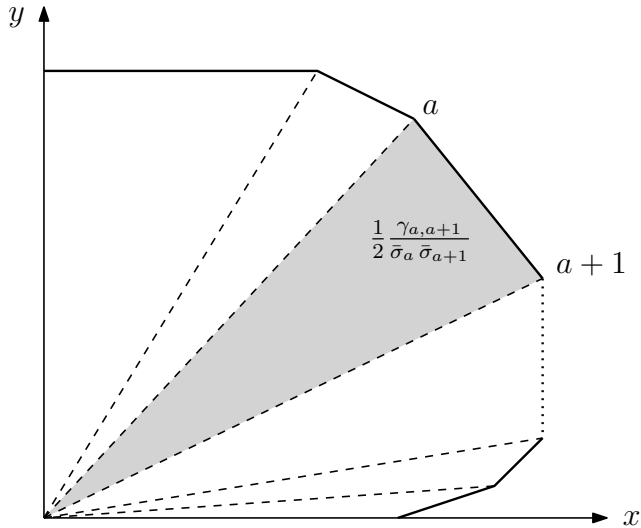

The exact free energy $F_{S^{3}}$ for these theories, which is a rational function of the CS levels $\{k_a\}$, was computed at large $N$ in \cite{Crichigno:2012sk} and we review the relevant results now.\footnote{The case of $\wh D_4$ was first studied in \cite{Gulotta:2011vp}.} Based on the explicit solution of the corresponding matrix models for various values of $n$, it was conjectured that for arbitrary $n\geq 4$, $F_{S^{3}}$  is determined by the area of a certain polygon $\mathcal P_n$ defined by the CS levels, which combined with \eqref{FS3vol} leads to a precise prediction for the volumes of the corresponding $Y_{7}$ manifolds, namely (the $n$--dependence of these manifolds will be made explicit in the next subsection)
\equ{\label{FS3}
\frac{\Vol(Y_{7})}{\Vol(S^{7})}=\frac{1}{4}\text{Area}(\mathcal P_n)\,,
}
where $\mathcal P_n$ is the polygon in $\bR^2$ defined by\footnote{This compact form of writing the polygon of  \cite{Crichigno:2012sk} is due to \cite{Moriyama:2015jsa}.}
\equ{\label{defP}
\mathcal P_n(x,y)=\left\{(x,y)\in \mathbb R^{2} \, \Big| ∑_{a=1}^n\(|y+p_a x|+|y-p_a x|\) -4|y|\leq 1\right\}·
}
Here $p$ is an $n$--dimensional vector such that at a given node $a$ the CS level is written as $k_{a}=\alpha_{(a)}\cdot p$ with $\alpha_{(a)}$ the root associated to that node. A typical polygon for generic values of CS levels is shown in Figure~\ref{Cone}. Writing $\text{Area}(\mathcal P_n)$ as the sum of the areas of the triangles defined by the origin and two consecutive vertices of the polygon, the AdS/CFT prediction \eqref{FS3} reads:
\equ{\frac{\text{Vol}(Y_{7})}{\text{Vol}(S^7)}=\frac{1}{2} \sum_{a=0}^{n} \frac{\gamma_{a,a+1}}{\bar \sigma_a\, \bar \sigma_{a+1}}\,,
\label{volPsigmas}
}
where  $\bar\sigma_{a} \equiv \sum_{b=1}^{n}\(|p_a-p_b|+|p_a+p_b|\)-4\, |p_a|$ for $a=1,\cdots,n$, and $\bar \sigma_0 = 2(n-2)\,, \bar \sigma_{n+1}=2 \sum_{b=1}^n |p_b|$. In addition $\gamma_{a,b}\equiv \left|  \beta_a \wedge \beta_b \right|$\footnote{Defining the wedge product $(a,b) \wedge (c,d) = (a d- b c)$.} with $\beta_a= (1, p_a)$ and $\beta_0=(0,1)$, $\beta_{n+1}=(1,0)$.

The physical meaning of $\mathcal P_n$ was clarified in \cite{Moriyama:2015jsa} (see also \cite{Assel:2015hsa}) where an elegant Fermi gas approach was used to study the matrix model at finite $N$, showing that $\mathcal P_n$ corresponds to the Fermi surface of the system at large $N$, and confirming the proposal for the free energy of  \cite{Crichigno:2012sk}. 

The goal for the rest of the paper is to derive \eqref{FS3} geometrically, by a direct  computation of $\Vol(Y_{7})$ using the localization method of section~\ref{NThyperkählerQ}. In order to do so, we must first identify the precise manifolds $Y_{7}$ dual to $\wh D$--quivers, which we do next.

\subsection{The Moduli Spaces}

The manifold $Y_{7}$ dual to a certain CS quiver gauge theory can be found by analyzing the moduli space of the field theory \cite{Aharony:2008ug,Jafferis:2008qz,Martelli:2008si}, which is obtained by setting the $D$--terms and $F$--terms to zero, and modding out by the appropriate gauge transformations. Thus, we need to specify the superpotential $W$ appearing in \eqref{action}. We can do this for a generic quiver. Consider a quiver with $n_V$ vertices corresponding to $U(n_aN)$ gauge groups (we assume all $n_a$ are coprime) and $n_E$ number of edges. Let us first set $N=1$. To determine the superpotential we follow the approach used in  \cite{Aharony:2008ug} by introducing an auxiliary chiral multiplet $Φ_{a}$ in the adjoint of the gauge group $a$ and superpotential  
$
W_{a}=-\frac{n_{a}k_{a}}{2}Φ_{a}^{2}+\sum_{i\to a}A_iΦ_{a}B_i\,;
$
here the sum is over all edges $i$ incident upon the node $a$ and $Φ_a=Φ_a^AT_A$, with $T_A$ the generators of the corresponding gauge group. To avoid cluttering the expressions we omit the gauge generators in what follows, but it should be clear where these sit. Since we will introduce a field $Φ_{a}$ for each node in the quiver it is convenient to introduce the notation $Φ\equiv(Φ_{1},\cdots,Φ_{n_V})^{\sT}$ and $AB\equiv(A_{1}B_{1},\cdots ,A_{n_E}B_{n_E})^{\sT}$ for nodes and edges, respectively. The full superpotential then reads $W=\sum_{a} W_{a} =-\frac{1}{2}Φ^{\sT} K Φ + Φ^{\sT} \I \, AB$, where $\I$ is the oriented incidence matrix of the quiver\footnote{This is defined to be a matrix which has a row for each vertex and column for each edge. The entry $\I_{ve}$ is $1$ if the edge $e$ comes into  vertex $v$, $-1$ if it comes out of it, and $0$ otherwise. These signs arise from the action of the group generators in  the terms $AΦ B\equiv A\,Φ^A\,T_A\, B$.} and $K$ is a diagonal matrix with entries  $K_{aa}=n_{a}k_{a}$. Since $Φ$ does not have a kinetic term it can be integrated out, leading to the superpotential
\equ{\label{Wgen}
W=\tfrac{1}{2} (AB)^{\sT}\, \I^{\sT}K^{-1}\I \, AB\,.
}

We are now in a position to determine the exact geometry of the moduli space for a general CS quiver. Varying $W$ with respect to $A$ and factoring out a $B$ gives the $F$--term equations $(AB)^{\sT}\, \I^{\sT}K^{-1}\I =0$. The $D$--term equations are obtained by simply replacing $AB→|A|^2-|B|^2$. Combining $A$ and $\bar{B}$ into a quaternion $q$, these three real equations combine into the hyperk\"ahler moment map equations $\sum_{j}Q^i_j(q_j^{†}(\s_\alpha)q_j)=0$, with $Q$ a charge matrix given by
\equ{\label{Qgen}
Q=\I^{\sT}K^{-1}\I\,.
}
This fully characterizes the quotient manifold for generic $\N=3$ quivers.\footnote{This is a slight generalization of the expression derived in section 2.5 of \cite{Jafferis:2008qz}, where we allow the gauge groups to have different ranks. } We now specialize this to  $\wh{D}_n$ quivers and begin with $\wh{D}_4$ for simplicity. 

\paragraph{$\bs{\wh{D}_4}$.}
Using the incidence matrix for $\wh D_{4}$ the superpotential \eqref{Wgen} reads 
\equ{W =\frac{1}{2}\left[\sum_{i=1}^{4}\frac{1}{k_{i}}\(A_{i}B_iA_{i}B_i\) +\frac{1}{2k_{5}}\Bigg(∑_{i=1}^4 B_i \cdot  A_{i}\Bigg)\Bigg(∑_{j=1}^4 B_j \cdot A_{j}\Bigg)\right],
}
where  $(A\cdot B)^{2}\equiv (A \sigma_AB)(A \sigma_AB)$ and  $\s_A=(I,\s_a)$. Varying $W$ with respect to $A_{i}$ gives the $F$--term equations:
\equ{\frac{1}{k_i}B_i(A_{i}B_i) +\frac{1}{2k_5}B_i(\s_A)∑_{j=1}^4\(B_j (\s_A)A_{j}\)=0\,.
\label{Ftermeq}}
Factoring out $B_i$, we have four matrix equations for each $i$. However, the $SU(2)$ part of the matrix gives the same equations for each $i$. In the quaternionic notation, all the $U(1)$ equations  $\big($from the $\s_0=I$ matrix in \eqref{Ftermeq}$\big)$ can be combined into the single equation
\equ{
∑_{j=1}^4 Q^i_j(q_j^{†}(\s_\alpha)q_j)=0 \qquad\text{with}\quad Q=\pmat{k_1+2k_5 & k_1 & k_1 & k_1 \\
k_2 & k_2+2k_5 & k_2 & k_2 \\  
k_3 & k_3 & k_3+2k_5 & k_3 \\
k_4 & k_4 & k_4 & k_4+2k_5}.
\label{Qmat4}}
Each column (lower index) in this matrix $Q$ represents a quaternion and each row (upper index) represents the $U(1)$ under which it is charged. This matrix can be obtained directly from \eqref{Qgen}; here we have multiplied each row `$i$' by $2k_{i}k_{5}$ for convenience, which amounts to an unimportant rescaling of the corresponding vector multiplets. We note that this matrix has only four rows although the original number of $U(1)$'s is five. The reason for this is that an overall diagonal $U(1)$ is decoupled as nothing is charged under it and so this row has been removed. In addition, imposing the  relation $k_1+k_2+k_3+k_4+2k_5=0$ one sees that rank$(Q)=3$ and hence another row must be removed (it does not matter which one)  to obtain the final charge matrix. We have thus shown that the moduli space is given by the hyperk\"ahler quotient $\bH^{8}///SU(2)×U(1)^3$ with the action of the group on the quaternions determined by the matrix in  \eqref{Qmat4}.

\paragraph{$\bs{\wh{D}_{n>4}}$.}
The extension to $\wh{D}_{n>4}$ quivers, with gauge group $U(2)^{n-3}×U(1)^4$, is direct. The superpotential \eqref{Wgen} can be written as:
\eqstn{W =\frac{1}{2}\Bigg[\sum_{i=1}^{4}\frac{1}{k_{i}}\(A_{i}B_iA_{i}B_i\)+\frac{1}{2k_{5}}\(A_{1}\cdot B_{1}+A_{2}\cdot B_{2}-A_{5}\cdot B_{5}\)^{2} \\
+\frac{1}{2k_{n+1}}\(A_{3}\cdot B_{3}+A_{4}\cdot B_{4}+A_{n}\cdot B_{n}\)^{2} +\sum_{a=6}^{n}\frac{1}{2k_{a}}\(A_{a-1}\cdot B_{a-1}-A_{a}\cdot B_{a}\)^{2}\Bigg].
}
Proceeding as above one concludes that the moduli space is given by the hyperk\"ahler quotient (at zero level)
\equ{\label{hyperkählerQM}
\C\(Y^{(n-3,n-1)}_7\)=\bH^{4n-8}///SU(2)^{n-3}×U(1)^{n-1} \,,
}
with the action of the group  on the quaternions given by the charge matrix (for $n>4$)  
\equ{Q=\pmat{k_1+2k_5 & k_1 & 0 & 0 & -k_1 & 0 & 0 & 0 \\
k_2 & k_2+2k_5 & 0 & 0 & -k_2 & 0 & 0 & 0 \\
0 & 0 & k_3+2k_{n+1} & k_3 & 0 & \cdots & \cdots_0 & k_3 \\
0 & 0 & k_4 & k_4+2k_{n+1} & 0 & \cdots & \cdots_0 & k_4 \\
-k_6 & -k_6 & 0 & 0 & k_5+k_6 & -k_5 & \ddots & 0 \\
0 & 0 & \vdots & \vdots &- k_7 & k_6+k_7 & \ddots_k & \ddots_0 \\
0 & 0 & \vdots_0 & \vdots_0 & \ddots & \ddots_k & \ddots_{k+k} & -k_{n-1} \\ 
0 & 0 & k_n & k_n & \ddots_0 & \ddots_0 & -k_{n+1} & k_n+k_{n+1}}.
\label{Qmatn}}
As above, the matrix is of rank $(n-1)$ after imposing $k_1+k_2+k_3+k_4+2(k_5+\cdots+k_{n+1})=0$ and one  (any) row should be removed. This matrix can be obtained directly from \eqref{Qgen}; here we have multiplied each row by the lowest common denominator of all the (nonzero) entries in that row for convenience.

We note that while the quaternionic dimension of the resulting spaces \eqref{hyperkählerQM} is two, there is only a single $U(1)$ remaining after the quotient and thus the spaces are non-toric. To see this, note that before gauging, the action for the $\wh D_n$ quiver has a $U(1)^n$ global symmetry, acting on each quaternion as $U(1)_i:(A_i,B_i)\to (e^{\i\theta}A_i, e^{-\i\theta}B_i)$ for $i=1,\cdots,n$. As shown above, the gauging removes $(n-1)$ of them, leaving a single $U(1)$ in the quotient manifold. This is also the case for the $\wh E$--quivers, as can be readily checked. For $\wh A $--quivers, in contrast, there is initially a $U(1)^n$ symmetry but the quotient removes only $(n-2)$ of them, hence the moduli spaces are toric.  

Since the moduli spaces are hyperk\"ahler quotients of the form \eqref{GenDnQuotient}, with $d=2,s=n-3,r=n-1$, we may apply the localization formula \eqref{intrepvol} to compute their volumes, which we do next.

\subsection{Volumes}

We are now in position to compute the volumes of tri-Sasaki Einstein manifolds dual to \mbox{$\wh D$--quivers.} For clarity of presentation, we sketch the basic steps for   $\wh D_{4}$ first and provide the details for general $\wh D_{n> 4}$ in Appendix~\ref{App:Dn computation}. Setting $d=2,s=1,r=3$ in \eqref{intrepvol} we have
\eqsn{\Vol\(Y_7^{(1,3)}\) &=\frac{32}{3}∫_0^∞dφ\, φ^2\(1+φ^2\)∫_{-∞}^∞d^3ϕ∏_{±,a=1}^4\frac{1}{1+\big(φ ±(Q_a^iϕ_i)\big)^2}\,·
}
To perform the $d^{3}\varphi$ integral it is convenient to use the Fourier transform identity
\equ{\frac{1}{[1+(φ+\varphi)^{2}][1+(φ-\varphi)^{2}]}=\frac14 \int\limits_{-\infty}^{\infty}dX\,\frac{e^{-|X|}}{2φ}\(\frac{e^{-\i |X| φ}}{φ-\i}+\frac{e^{\i |X| φ}}{φ+\i}\)e^{\i\varphi X}
}
for each term in $∏_{a=1}^4$. Performing the $d^{3}\varphi$ integrals  generates $(2π)^3δ^3({\textstyle ∑_a}Q^i_aX_a)$,\footnote{As explained in \cite{Gulotta:2011si}, for non-coprime entries in the charge matrix $Q$ there is an extra numerical factor dividing the $δ$--functions. But in that case, $\Vol\(U(1)^r\)$ is also not simply $(2π)^r$ but needs to be divided by the same factor, so the result being derived here is valid for generic $Q$.} which can be integrated away by writing $X_a=k_a x$; it is directly checked from \eqref{Qmat4} that $∑_aQ_a^ik_a=0$. Thus, we obtain
\equ{\Vol\(Y_7^{(1,3)}\)=\frac{π^3}{3}∫_{-∞}^∞dx∫_0^∞dφ\,\frac{\exp{-∑_a|k_a x|}}{φ^2(1+φ^2)^3}∏_{a=1}^4 \left[\frac{1}{2}\(e^{-\i |k_ax| φ}(φ+\i)+e^{\i |k_ax| φ}(φ-\i)\)\right].
\label{volD4xfrem}
}
We now perform the $φ$ integral by residues, converting $∫_0^∞dφ →\frac{1}{2}∫_{-∞}^∞dφ$ as the integrand is an even function of $φ$. We note that expanding the product of exponentials in \eqref{volD4xfrem} gives a total of sixteen terms and the precise integration contour in the complex plane needs to be chosen separately for each one. This is  because the coefficient of $\i φ|x|$ in each term can be any one of the combinations $±|k_1|±|k_2|±|k_3|±|k_4|$. Thus, in order to decide how to close the contour at $∞$, we choose a particular ordering of $k$'s. It is convenient to go to the basis $k_a → α_{(a)}.p$ and order the $p$'s according to  $p_1≥p_2≥p_3≥p_4≥0$ (this is simply a choice and one should repeat this for all possible orderings). This results in
\eqst{\Vol\(Y_7^{(1,3)}\)=\frac{π^4}{3}\frac{1}{2}∫_{-∞}^∞dx\,\exp{-2\(p_1+p_3\)|x|}\left[-\frac{1}{8}\exp{-2\(p_1 -p_3\)|x|} +\frac{1}{2}\exp{-2(p_2-p_3)|x|}\(1+(p_2-p_3)|x|\) \right. \\ \nonumber
\left. -\frac{1}{8}\(\exp{-2(p_2 -p_4)|x|}+\exp{-2(p_2 +p_4)|x|}\) \right].
}
Finally, integrating over $x$ gives  
\eqs{
\frac{\Vol\(Y_7^{(1,3)}\)}{\Vol\(S^7\)}&=-\frac{1}{32p_1} +\frac{2p_1 +3p_2-p_3}{8\(p_1+p_2\)^2} -\frac{1}{16(p_1+p_2+p_3 -p_4)} -\frac{1}{16\sum_{b=1}^4 p_b}\, \nn
&=\frac{1}{2}∑_{a=0}^4\frac{γ_{a,a+1}}{\bar{\s}_a\bar{\s}_{a+1}} =\frac{1}{4}\text{Area}\({\mathcal P}_4\)\,,
}
where in the second line we used the definitions below \eqref{volPsigmas} and the ordering of $p$'s we have chosen (one may check that the last line above gives the result of the integral for all possible orderings). Thus, we have shown that for $n=4$ one exactly reproduces the field theory prediction \eqref{FS3}.

For generic $n\geq4$ the volume formula reads
\eqss{\frac{\Vol\(Y_7^{(n-3,n-1)}\)}{\Vol\(S^7\)} &=\frac{4^{2n-5}}{(π^2)^{n-3}(2π)^{n-1}}∏_{i=1}^{n-3}∫_0^∞ dφ_i(4πφ_i^2)\big(1+φ_i^2\big)∏_{j=1}^{n-1}∫_{-∞}^∞ dϕ_j \\
&\quad ×∏_{±,a=1}^2\frac{1}{1+\big(φ_1±Q_a^iϕ_i\big)^2} ∏_{±,a=3}^4\frac{1}{1+\big(φ_{n-3}^2±Q_a^iϕ_i\big)^2} \\
&\quad ×∏_{±,a=5}^n\frac{1}{\(1+(φ_{a-4} ±φ_{a-3} ±Q_a^iϕ_i)^2\)}\,·
}
The integrals can be performed by the same steps as in the  $\wh D_4$ case.  Assuming the ordering $p_1≥p_2≥\cdots≥p_n≥0$ one finds (see Appendix~\ref{App:Dn computation} for details):
\eqs{\frac{\Vol\(Y_7^{(n-3,n-1)}\)}{\Vol\(S^7\)} &=\frac{1}{16}\sum_{a=1}^{n-3}\frac{c_a}{\sum_{b=1}^{a-1}p_b +(n-a-1)p_{a}}  + \frac{2\sum_{b=1}^{n-3} p_b +3 p_{n-2}-p_{n-1}}{8\(\sum_{b=1}^{n-2} p_b\)^2} \nn
&\qquad\quad -\frac{1}{16}\(\frac{1}{\sum_{b=1}^{n-1}p_b -p_n}+\frac{1}{\sum_{b=1}^n p_b}\)\nn
&=\frac{1}{4}\text{Area}(\mathcal P_n)\,,
\label{volDnloc}
}
in perfect agreement with the field theory prediction \eqref{FS3}!

\section{Summary and Outlook}

This paper contains two main results. The first is an explicit integral formula  computing the volumes of tri-Sasaki Einstein manifolds given by nonabelian hyperk\"ahler quotients. This is a generalization of the formula derived by Yee \cite{Yee:2006ba} in the Abelian case.  The second result concerns the study of 3d $\N=3$ CS matter theories. We identified  the precise (non-toric) tri-Sasaki Einstein manifolds describing the gravity duals of $\wh D$--quivers and computed their volumes, showing perfect agreement with the field theory prediction of \cite{Crichigno:2012sk}. This greatly expands the detailed tests of AdS$_{4}$/CFT$_{3}$ available for non-toric cases. 

One may also consider CS  $\wh E$--quivers, whose free energies were computed in \cite{Crichigno:2012sk}. In this case the corresponding hyperk\"ahler quotients are $\wh E_6:\, \mathbb H^{24}///SU(3)\times SU(2)^3\times U(1)^5$, $\wh E_7:\, \mathbb H^{48}///SU(4)\times SU(3)^2\times SU(2)^3\times U(1)^6$, and $\wh E_8:\,\mathbb H^{120}///SU(6)\times SU(5)\times SU(4)^2\times SU(3)^2\times SU(2)^2\times U(1)^7$. The volume integrals can be written using  \eqref{genvolSUNU1} and the relevant charge matrices  \eqref{Qgen}. Although we have not computed these integrals explicitly one should be able to do so with the same techniques used here for $\widehat D$--quivers.  An  open question regarding $\widehat E$--quivers is whether  they  admit a Fermi gas description, along the lines of \cite{Marino:2011eh} for  $\wh A$--quivers  and \cite{Moriyama:2015jsa,Assel:2015hsa} for $\wh D$--quivers. If so, the integral volume formula may elucidate the form of the Fermi surface in the large $N$ limit.

The localization approach can also be applied to nonabelian K\"ahler quotients. This is the relevant setting for AdS$_{5}$/CFT$_{4}$, where few non-toric examples are known.  An important distinction, however, is that the quotient ensures only the K\"ahler class of the quotient manifold and not its metric structure. In this case one would have to combine this approach with the principle of volume minimization, along the lines of \cite{Martelli:2006yb,Martelli:2005tp}. It is our hope that the formulas presented here will also be valuable in this context.

Finally, one may also consider quivers whose nodes represent $SO(N)$ or $USp(2N)$ gauge groups. Related to this, it may be interesting to consider the interplay of the volume formulas with the folding/unfolding procedure of  \cite{Gulotta:2012yd}.

\section*{\centering Acknowledgements}
DJ thanks Kazuo Hosomichi for many insightful discussions on the topic of localization. We are also grateful to Kazuo and Chris Herzog for suggestions and comments on the manuscript. DJ is supported in most part by MOST grant no. 104-2811-M-002-026. PMC is supported by Nederlandse Organisatie voor Wetenschappelijk Onderzoek (NWO) via a Vidi grant. The work of PMC is part of the Delta ITP consortium, a program of the NWO that is funded by the Dutch Ministry of Education, Culture and Science (OCW). PMC would like to thank NTU and ITP at Stanford University for kind hospitality where part of this work was carried out.

\appendix
\section[\texorpdfstring{$\wh D_{n}$ CS Quivers}{Dn CS Quivers}]{$\bs{\wh D_{n}}$ CS Quivers}\label{App:Dn computation}

Here we provide the details leading to the main result for CS $\wh{D}_n$ quivers \eqref{volDnloc}. For generic $n$ the volume formula reads:
\eqs{\Vol\(Y_7^{(n-3,n-1)}\) &=\(\frac{π^4}{3}\) \frac{4^{2n-5}}{(π^2)^{n-3}(2π)^{n-1}}∏_{i=1}^{n-3}∫_0^∞ dφ_i(4πφ_i^2)\big(1+φ_i^2\big)∏_{j=1}^{n-1}∫_{-∞}^∞ dϕ_j \nn
&\quad ×∏_{±,a=1}^2\frac{1}{1+\big(φ_1±Q_a^iϕ_i\big)^2} ∏_{±,a=3}^4\frac{1}{1+\big(φ_{n-3}^2±Q_a^iϕ_i\big)^2} \nn
&\quad ×∏_{±,a=5}^n\frac{1}{\(1+(φ_{a-4} ±φ_{a-3} ±Q_a^iϕ_i)^2\)}\,·
}
The basic procedure follows the same logic of the $\wh D_{4}$ case. We first exponentiate the denominators by introducing some $∫dy_a$'s, perform the $ϕ$--integrals to generate $δ(∑_aQ^i_ay_a)$--functions, and solve the equations $∑_aQ^i_ay_a=0$ by $y_a=κ_a x$ such that $∑_aQ^i_aκ_a=0$ where $κ_a=\{p_1+p_2, p_1-p_2, p_{n-1}-p_n, p_{n-1}+p_n, 2p_3, 2p_4, ⋯, 2p_{n-2}\}$ (up to some signs but since only $|κ_a|$ are needed below these are not important). Now, assuming 
\equ{\label{orderingpapp}
p_1≥p_2≥\cdots≥p_n≥0\,,
}
all  $κ_a\geq0$ and thus we may replace $|κ_a|\toκ_a$. Next, we perform all the $y_a$ integrals obtaining
\eqst{\frac{\Vol\(Y_7^{(n-3,n-1)}\)}{\Vol\(S^7\)} =\frac{4^{2n-5}4^{n-3}}{π^{n-3}}\frac{1}{4^4}\frac{1}{32^{n-4}}∫_{-∞}^∞ dx\,\exp{-∑_{a=1}^nκ_a|x|} ∏_{i=1}^{n-3}∫_0^∞ dφ_i\,φ_i^2\big(1+φ_i^2\big) \\
×∏_{a=1}^2\D_{κ_a}(φ_1,x)∏_{a=3}^4\D_{κ_a}(φ_{n-3},x) ∏_{a=5}^n\D_{κ_a}(φ_{a-4},φ_{a-3},x)\,,
}
where
\eqsg{\D_{κ_a}(φ_i,x) &=\frac{φ_i\cos(φ_i κ_a |x|)+\sin(φ_i κ_a|x|)}{φ_i(1+φ_i^2)} \\
\D_{κ_a}(φ_i,φ_j,x) &=\frac{\left[\begin{array}{c}φ_iφ_j(φ_i^2-φ_j^2)(5+φ_i^2+φ_j^2)\cos(φ_i κ_a|x|)\cos(φ_j κ_a|x|) \\ +2(1+φ_i^2)(1+φ_j^2)(φ_i^2-φ_j^2)\sin(φ_i κ_a|x|)\sin(φ_j κ_a|x|) \\ -φ_i(1+φ_i^2)(1-φ_i^2+5φ_j^2)\cos(φ_i κ_a|x|)\sin(φ_j κ_a|x|) \\ +φ_j(1+φ_j^2)(1+5φ_i^2-φ_j^2)\sin(φ_i κ_a|x|)\cos(φ_j κ_a|x|) \end{array}\right]}{φ_iφ_j(φ_i^2-φ_j^2)(1+φ_i^2)(1+φ_j^2)(1+(φ_i+φ_j)^2)(1+(φ_i -φ_j)^2)} \,·
}
By performing the integrals in decreasing order of $φ$'s, starting from $φ_{n-3},⋯, φ_1$ a pattern emerges. Here are a few intermediate steps:
\eqst{\I_{n-3}=∫dφ_{n-3}\,φ_{n-3}^2\big(1+φ_{n-3}^2\big)∏_{a=3}^4\D_{κ_a}(φ_{n-3},x) \D_{κ_n}(φ_{n-4},φ_{n-3},x) =\frac{π}{2}∏_{a=3}^4\D_{κ_a}(φ_{n-4},x) \\
+\frac{π}{4}\exp{2(-p_{n-2}+p_{n-1})|x|}\frac{φ_{n-4}(φ_{n-4}^2 -5)\cos(φ_{n-4}κ_n|x|)+2(2φ_{n-4}^2 -1)\sin(φ_{n-4}κ_n|x|)}{φ_{n-4}(1+φ_{n-4}^2)^2(4+φ_{n-4}^2)}\,·
\label{firstphi}}
Let us define another $\D$ to keep the expressions relatively compact:
\equ{\D_{κ_a}(φ_i,x;\fm) =\frac{φ_i\(φ_i^2 -\fm(\fm+1) +1\)\cos(φ_i κ_a|x|) +\fm(2φ_i^2 -\fm +1)\sin(φ_i κ_n|x|)}{φ_i\(1+φ_i^2\)\((\fm-1)^2+φ_i^2\)\(\fm^2+φ_i^2\)}\,·
}
Thus the relevant expression in \eqref{firstphi} can be labelled $\D_{κ_n}(φ_{n-4},x;2)$. Proceeding further with the integrals we have
\eqst{\I_{n-3>j>1} =∫dφ_j\,φ_j^2\big(1+φ_j^2\big)\D_{κ_{j+3}}(φ_{j-1},φ_j,x)\,\I_{j+1} \\
=\(\frac{π}{2}\)^{n-2-j}\left[∏_{a=3}^4\D_{κ_a}(φ_{j-1},x) +\frac{1}{2}∑^{a=n}_{j+3}\exp{-2\((n-a+1)p_{a-2}-∑_{b=a-1}^{n-1}p_b\)|x|}\D_{κ_a}(φ_{j-1},x;n-a+2)\right].
}
The final $φ_1$--integral then gives
\eqs{\I_1 &=∫dφ_1\,φ_1^2\big(1+φ_1^2\big)∏_{a=1}^2\D_{κ_a}(φ_1,x)\,\I_2 \nn
&=\(\frac{π}{2}\)^{n-3}\left[\frac{c_1}{8}\exp{-2\((n-3)p_1 -∑_{b=3}^{n-1}p_b\)|x|} +∑_{a=2}^{n-3}\frac{c_a}{8}\exp{-2\(∑_{b=2}^{a-1}p_{b} +(n-a-1)p_{a} -∑_{b=3}^{n-1}p_b\)|x|} \right. \nn
&\quad\left. +\frac{1}{2}\exp{-2(p_2-p_{n-1})|x|}\(1+(p_{n-2}-p_{n-1})|x|\) -\frac{1}{8}\(\exp{-2(p_2 -p_n)|x|}+\exp{-2(p_2 +p_n)|x|}\) \right],
}
where $c_a=\frac{-2}{(n-a-1)(n-a-2)}$. This expression is also valid for $\wh{D}_4$, as can be easily checked.

Finally, performing the integral over $x$ gives
\eqs{\frac{\Vol\(Y_7^{(n-3,n-1)}\)}{\Vol\(S^7\)} &= \frac{2^{n-4}}{π^{n-3}}∫_{-∞}^∞ dx\,\exp{-2\(p_1+∑_{b=3}^{n-1}p_b\)|x|}\;\I_1 \label{intgx}\nn
&=\frac{1}{16}\sum_{a=1}^{n-3}\frac{c_a}{\sum_{b=1}^{a-1}p_b +(n-a-1)p_{a}}  + \frac{2\sum_{b=1}^{n-3} p_b +3 p_{n-2}-p_{n-1}}{8\(\sum_{b=1}^{n-2} p_b\)^2} \nn
&\quad -\frac{1}{16}\(\frac{1}{\sum_{b=1}^{n-1}p_b -p_n}+\frac{1}{\sum_{b=1}^n p_b}\)·
}
The expression appearing on the right hand side of \eqref{intgx} is precisely the area of the polygon \eqref{defP} (see \cite{Crichigno:2012sk} for details). Indeed, using the definitions below \eqref{volPsigmas} and the ordering \eqref{orderingpapp}, this becomes
\equn{\frac{\Vol\(Y_7^{(n-3,n-1)}\)}{\Vol\(S^7\)} =\frac{1}{2}∑_{a=0}^n\frac{γ_{a,a+1}}{\bar{\s}_a\bar{\s}_{a+1}} =\frac{1}{4}\text{Area}({\mathcal P}_n) \,,
}
as we wanted to show.

\section{Other Examples}\label{AVolofLRs}

In this Appendix we provide other examples of applications of the formula \eqref{intrepvol}.

\subsection{A Lindström-Roček Space}

Consider a Lindström-Roček Space \cite{Lindstrom:1983rt} given by the hyperk\"ahler quotient $\bH^6///U(2)$. 
This amounts to setting  $d=2,s=1,r=1$ in \eqref{intrepvol} and the volume reads
\eqsn{\Vol\(Y^{(1,1)}_7\) &=\frac{32π^4}{6[(π^2)(2π)]}∫_0^{∞}dφ (4π φ^2)\(1+φ^2\)∫_{-∞}^∞dϕ \frac{1}{\(1+2\big(φ^2 +ϕ^2\big) +\big(φ^2 -ϕ^2\big)^2\)^3} \nn
&=\frac{32π^2}{3}∫_0^{∞}dφ\,φ^2\(1+φ^2\)\left[\frac{3π(21+6φ^2+φ^4)}{256\(1+φ^2\)^5}\right] \nn
&=\frac{π^3}{8}∫_0^{∞}dφ\,\frac{φ^2(21+6φ^2+φ^4)}{\(1+φ^2\)^4} =\frac{π^4}{8}\,·
}
One can  verify that this is the correct value by explicit construction of the hyperk\"ahler potential. Following \cite{Lindstrom:1983rt}, the hyperkähler cone $\bH^6///U(2)$ is described by the following action (with all FI parameters vanishing)
\equ{S=∫d^8z\left[\bar{Φ}_{a+}^m\big(e^V\big)^a_bΦ^b_{m+} +Φ^m_{a-}\big(e^{-V}\big)^a_b\bar{Φ}^b_{m-}\right] +\left[∫d^6zΦ^b_{m+}S_b^aΦ^m_{a-} +h.c.\right].
\label{LRaction}}
Here, $m=1,2,3$ and $a=1,2$ is the $U(2)$ index. This gives the following equations of motion
\eqs{Φ^b_{m+}\bar{Φ}^m_{a+}\big(e^V\big)^a_b -\big(e^{-V}\big)^a_b\bar{Φ}^b_{m-}Φ^m_{a-} &=0\\
Φ^b_{m+}Φ^m_{a-} &=0\,.
}
Solving the latter equation by
\eqs{Φ^a_+ &=\(K^a_+, \frac{\i K_{1-}K^a_+}{\sqrt{K^a_+K_{a-}}}, \frac{\i K_{2-}K^a_+}{\sqrt{K^a_+K_{a-}}}\) \\
Φ_{a-} &=\(K_{a-}, \frac{\i K_{a-}K^1_+}{\sqrt{K^a_+K_{a-}}}, \frac{\i K_{a-}K^2_+}{\sqrt{K^a_+K_{a-}}}\)^{\!\!\sT},
}
where we have chosen a particular gauge, and plugging the solution for $e^V$ back in \eqref{LRaction} leads to the action
\eqs{S &=\Tr∫d^8z\sqrt{4Φ^b_{m+}\bar{Φ}^m_{c+} \bar{Φ}^c_{m-}Φ^m_{a-}} \nn
&=2∫d^8z\sqrt{\(K^1_+\bar{K}_{1+} +K^2_+\bar{K}_{2+} +κ\)\(K_{1-}\bar{K}^1_- +K_{2-}\bar{K}^2_- +κ\)}\,,
}
where $κ^2=(K^1_+K_{1-}+K^2_+K_{2-})(\bar{K}^1_+\bar{K}_{1-}+\bar{K}^2_+\bar{K}_{2-})$. The metric is given by $g_{i\bar{j}}=∂_{i\bar{j}}\K$ where Kähler potential $\K$ is obtained from $S=∫d^8z \K$. It turns out that
\equ{g≡\det g_{i\bar{j}}=2^8.
}

We use the following coordinate transformation to spherical polar coordinates
\eqsg{K^1_+ &=r \cos χ \cos\tfrac{θ_1}{2}e^{\frac{\i}{2}(-φ_1 +2ψ_1)} \\
K_{1-} &=r \cos χ \sin\tfrac{θ_1}{2}e^{\frac{\i}{2}(-φ_1 -2ψ_1)} \\
K^2_+ &=r \sin χ \cos\tfrac{θ_2}{2}e^{\frac{\i}{2}(-φ_2+2ψ_2)} \\
K_{2-} &=r \sin χ \sin\tfrac{θ_2}{2}e^{\frac{\i}{2}(-φ_2-2ψ_2)} \,.
}
Here, $r$ is the radial coordinate and $θ_i,φ_i,ψ_i$ are the usual 3D spherical coordinates so $θ_i\in[0,π]$, $φ_i\in[0,2π)$ and $ψ_i\in[0,2π)$. The limit of $χ\in[0,\frac{π}{2}]$ is chosen such that the `flat' action gives flat metric on $R_+×S⁷$. The determinant of the Jacobian of this transformation is
\equ{J_s=r^7 \cos^3χ \sin^3χ \sin θ_1 \sin θ_2.
}
In these coordinates the metric is not explicitly conical  (there are off-diagonal terms between $dr$ and spherical coordinates) but $g_{rr}$ is a complicated function of spherical coordinates only and rescaling $r→\frac{ρ}{\sqrt{g_{rr}}}$ one obtains  the conical metric $dρ^2+ρ^2dΩ_7^2$. The determinant of this radial transformation is
\equ{J_r=\frac{1}{\sqrt{g_{rr}}}\,·
}
Combining all the above determinants, taking square root and (numerically) integrating over the spherical coordinates gives us the volume of the seven-dimensional base of the hyperkähler cone:
\equ{\Vol(Ω_7)=∫_{Ω_7}\sqrt{g}\,J_s\,J_r|_{ρ→1} =∫_{Ω_7}\frac{16\cos^3χ \sin^3χ \sin θ_1 \sin θ_2}{g_{rr}^4} =\frac{π^4}{8}\,·
}

\subsection[\texorpdfstring{Volume of $\bD$--orbifolded $S^3$}{Volume of D-orbifolded S³}]{Volume of $\pmb{\bD}$--orbifolded $\bs{S^3}$}\label{BVolS3D}

Here we provide details of the calculation for ALE instantons of section~\ref{sec:ALE instantons} for generic $\mathbb D_{k-2}$. The volume integral reads:
\eqs{\Vol\(Y^{(k-3,k)}_3\) &=\frac{2^{3(k-3)+k+1}π^2}{(π^2)^{k-3}×(2π)^k}∫∏_{i=1}^{k-3} dφ_i\(4πφ_i^2\)\(1+φ_i^2\) ∏_{j=1}^k dϕ_j\,∏_{±}\frac{1}{1+\(φ_1 ±ϕ_4\)^2} \nn
&\quad ×∏_{±}\frac{1}{1+\(φ_1 ±(ϕ_1+ϕ_4)\)^2} ∏_{±,a=2}^3\frac{1}{1+\(φ_{k-3} ±(ϕ_a+ϕ_k)\)^2} \nn
&\quad ×∏_{±,a=1}^{k-4}\frac{1}{1+\((φ_{a}±φ_{a+1}) ±(ϕ_{a+3}-ϕ_{a+4})\)^2}\,·
}
Using Fourier transform to exponentiate all the denominators, we obtain
\eqsn{\Vol\(Y^{(k-3,k)}_3\) &=\frac{2^{5k-14}}{π^{2k-5}}∫∏_{i=1}^{k-3} dφ_i\, φ_i^2\(1+φ_i^2\) ∏_{j=1}^k dϕ_j ∏_{±,a=1}^4 dy_a^± ∏_{±,b=1}^{k-4}dη_b^±d\bar{η}_b^± \nn
&\quad ×\(\half\)^8\exp{-∑_{±,a=1}^4|y_a^±|+\i∑_±\(y_1^±(φ_1 ±ϕ_4) +y_2^±\(φ_1 ±(ϕ_1+ϕ_4)\) +∑_{a=2}^3 y_{a+1}^±\(φ_{k-3} ±(ϕ_a+ϕ_k)\)\)} \nn
&\quad ×\(\half\)^{4(k-4)}\exp{∑_{±,b=1}^{k-4}\(\!-|η_b^±|+\i η_b^±\(\!(φ_b+φ_{b+1}) ±(ϕ_{b+3}-ϕ_{b+4})\!\) -|\bar{η}_b^±|+\i \bar{η}_b^±\(\!(φ_b-φ_{b+1}) ±(ϕ_{b+3}-ϕ_{b+4})\!\)\!\)} \nn
&=\frac{2^{k-6}}{π^{2k-5}}∫∏_{i=1}^{k-3} dφ_i\, φ_i^2\(1+φ_i^2\) ∏_{j=1}^k dϕ_j ∏_{±,a=1}^4 dy_a^± ∏_{±,b=1}^{k-4}dη_b^±d\bar{η}_b^± \nn
&\quad ×\exp{-∑_{±,a=1}^4|y_a^±|-∑_{±,b=1}^{k-4}\(|η_b^±|+|\bar{η}_b^±|\)}\,\exp{\i∑_±\(φ_1\(y_1^± +y_2^±\) +φ_{k-3}\(y_3^± +y_4^±\)\)} \nn
&\quad ×\exp{\i∑_{a=1}^3ϕ_a\(y_{a+1}^+ -y_{a+1}^-\) +\i ϕ_4\(y_1^+ -y_1^- +y_2^+ -y_2^-\) +\i ϕ_k\(y_3^+ -y_3^- +y_4^+ -y_4^-\)} \nn
&\quad ×\exp{\i φ_1\(η_1^++η_1^- +\bar{η}_1^+ +\bar{η}_1^-\) +\i φ_{k-3}\(η_{k-4}^+ +η_{k-4}^- -\bar{η}_{k-4}^+ -\bar{η}_{k-4}^-\) +\i∑_{±,b=2}^{k-4}φ_b\(η_{b-1}^± -\bar{η}_{b-1}^± +η_b^± +\bar{η}_b^±\)} \nn
&\quad ×\exp{\i ϕ_4\(η_1^+ -η_1^- +\bar{η}_1^+ -\bar{η}_1^-\) -\i ϕ_k\(η_{k-4}^+ -η_{k-4}^- +\bar{η}_{k-4}^+ -\bar{η}_{k-4}^-\) +\i∑_{±,b=2}^{k-4}ϕ_{b+3}\(∓η_{b-1}^± ∓\bar{η}_{b-1}^± ±η_b^± ±\bar{η}_b^±\)}\,.
}
We can perform the three $ϕ_a, a=1,2,3$ integrals to generate three $δ$--functions, which can be used to do $y_a^-, a=2,3,4$ integrals as follows:
\eqsn{\Vol\(Y^{(k-3,k)}_3\) &=\frac{2^{k-6}}{π^{2k-5}}∫∏_{i=1}^{k-3} dφ_i\, φ_i^2\(1+φ_i^2\) ∏_{j=4}^k dϕ_j ∏_{a=1}^4 dy_a^+dy_1^- ∏_{±,b=1}^{k-4}dη_b^±d\bar{η}_b^± \nn
&\quad ×\exp{-∑_±|y_1^±| -2∑_{a=2}^4|y_a^+|-∑_{±,b=1}^{k-4}\(|η_b^±|+|\bar{η}_b^±|\)}\,\exp{\i\(φ_1\(y_1^+ +y_1^- +2y_2^+\) +2φ_{k-3}\(y_3^+ +y_4^+\)\)} \nn
&\quad ×\exp{\i φ_1\(η_1^++η_1^- +\bar{η}_1^+ +\bar{η}_1^-\) +\i φ_{k-3}\(η_{k-4}^+ +η_{k-4}^- -\bar{η}_{k-4}^+ -\bar{η}_{k-4}^-\) +\i∑_{±,b=2}^{k-4}φ_b\(η_{b-1}^± -\bar{η}_{b-1}^± +η_b^± +\bar{η}_b^±\)} \nn
&\quad ×(2π)^3\exp{\i ϕ_4\(y_1^+ -y_1^-\)}\exp{\i ϕ_4\(η_1^+ -η_1^- +\bar{η}_1^+ -\bar{η}_1^-\) -\i ϕ_k\(η_{k-4}^+ -η_{k-4}^- +\bar{η}_{k-4}^+ -\bar{η}_{k-4}^-\)} \nn
&\quad ×\exp{\i∑_{±,b=2}^{k-4}ϕ_{b+3}\(∓η_{b-1}^± ∓\bar{η}_{b-1}^± ±η_b^± ±\bar{η}_b^±\)}\,.
}
This form now shows that all the remaining $ϕ$--integrals can be done similarly to generate more $δ$--functions involving $η$'s and then all the remaining $y^+$ and $η^±$-integrals can be performed, leaving only the $φ$--integrals.
\eqsn{\Vol\(Y^{(k-3,k)}_3\) &=\frac{2^{k-6}}{π^{2k-5}}(2π)^k∫∏_{i=1}^{k-3} dφ_i\, φ_i^2\(1+φ_i^2\) ∏_{a=1}^4 dy_a^+ ∏_{±,b=1}^{k-4}dη_b^±d\bar{η}_b^±\, \exp{-2∑_{a=1}^4|y_a^+|} \nn
&\quad ×\exp{2\i\(φ_1\(y_1^+ +y_2^+\) +φ_{k-3}\(y_3^+ +y_4^+\)\)} ∏_{b=1}^{k-4}δ\(η_b^+ -η_b^- +\bar{η}_b^+ -\bar{η}_b^-\) \exp{-∑_{±,b=1}^{k-4}\(|η_b^±|+|\bar{η}_b^±|\)} \nn
&\quad ×\exp{\i φ_1\(η_1^++η_1^- +\bar{η}_1^+ +\bar{η}_1^-\) +\i φ_{k-3}\(η_{k-4}^+ +η_{k-4}^- -\bar{η}_{k-4}^+ -\bar{η}_{k-4}^-\) +\i∑_{±,b=2}^{k-4}φ_b\(η_{b-1}^± -\bar{η}_{b-1}^± +η_b^± +\bar{η}_b^±\)} \nn
&=\frac{2^{2k-6}}{π^{k-5}}∫∏_{i=1}^{k-3} dφ_i\,\frac{φ_i^2\(1+φ_i^2\)}{\(1+φ_1^2\)^2\(1+φ_{k-3}^2\)^2} ∏_{±,b=1}^{k-4}dη_b^±d\bar{η}_b^+ \nn
&\quad ×\exp{-∑_{b=1}^{k-4}\(|η_b^+|+|η_b^-|+|\bar{η}_b^+|+|η_b^+-η_b^-+\bar{η}_b^+|\)} \nn
&\quad ×\exp{2\i φ_1\(η_1^+ +\bar{η}_1^+\) +2\i φ_{k-3}\(η_{k-4}^- -\bar{η}_{k-4}^+\) +2\i∑_{b=2}^{k-4}φ_b\(η_{b-1}^- -\bar{η}_{b-1}^+ +η_b^+ +\bar{η}_b^+\)} \nn
&=\frac{2^{2k-6}}{π^{k-5}}∫∏_{i=1}^{k-3} dφ_i\, \frac{φ_i^2\(1+φ_i^2\)}{\(1+φ_1^2\)^2\(1+φ_{k-3}^2\)^2} \nn
&\quad ×∏_{b=1}^{k-4}\frac{5+φ_b^2+φ_{b+1}^2}{2(1+φ_b^2)(1+φ_{b+1}^2)(1+(φ_b+φ_{b+1})^2)(1+(φ_b-φ_{b+1})^2)} \nn
&=\frac{2^{k-2}}{π^{k-5}}∫∏_{i=1}^{k-3} dφ_i\,\frac{φ_i^2}{(1+φ_1^2)^2(1+φ_{k-3}^2)} \nn
&\quad ×∏_{b=1}^{k-4}\frac{5+φ_b^2+φ_{b+1}^2}{(1+φ_{b+1}^2)(1+(φ_b+φ_{b+1})^2)(1+(φ_b-φ_{b+1})^2)}\,·
}
Finally, performing all the $φ$--integrals one-by-one with the residue algorithm used in the main text (and  Appendix \ref{App:Dn computation}), we get
\eqs{\Vol\(Y^{(k-3,k)}_3\) &=\frac{{2^{k-2}}}{π^{k-5}}\frac{π}{{4}\(1+(k-3)\)^2}∏_{a=1}^{k-4}\frac{(a+2)π}{{2}(a+1)} \nn
&=\frac{π^2}{2(k-2)}\,,
}
as expected.

\bibliographystyle{utphys}
\bibliography{References} 

\end{document}